\documentclass[preprint,5p,twocolumn,10pt]{elsarticle}

\usepackage[pdftex,hypertexnames=false,linktocpage=true]{hyperref}
\hypersetup{colorlinks=true,linkcolor=red,anchorcolor=darkgreen,citecolor=green!50!blue,filecolor=darkgreen,urlcolor=green,bookmarksnumbered=true,pdfview=FitB}

\usepackage{lipsum}
\usepackage{mathtools,cuted}

\usepackage{amsmath}
 \usepackage{widetext}

\usepackage{amssymb}
\usepackage{graphicx}
\usepackage{multirow}

\usepackage{array}
\usepackage{booktabs}
\usepackage{arydshln}
\usepackage{multirow}

\usepackage{braket}

\graphicspath{{../../plots/}}

\begin{document}

\begin{frontmatter}
\title{Quark and gluon distributions in $\rho$-meson from basis light-front quantization}

\author[a,b,c]{Satvir Kaur}
\ead{satvir@impcas.ac.cn}
\author[a,b,c]{Jiatong Wu}
\ead{wujt@impcas.ac.cn}
\author[a,d]{Zhi Hu\corref{cor1}}
\ead{huzhi0826@gmail.com}

\author[a,b,c]{Jiangshan Lan\corref{cor1}}
\ead{jiangshanlan@impcas.ac.cn}

\author[a,b,c]{Chandan Mondal\corref{cor1}}
\ead{mondal@impcas.ac.cn}

\author[a,b,c]{Xingbo Zhao}
\ead{xbzhao@impcas.ac.cn}

\author[xyz]{James~P.~Vary}
\ead{jvary@iastate.edu}

\address[a]{Institute of Modern Physics, Chinese Academy of Sciences, Lanzhou 730000, China}
\address[b]{School of Nuclear Science and Technology, University of Chinese Academy of Sciences, Beijing 100049, China}
\address[c]{CAS Key Laboratory of High Precision Nuclear Spectroscopy, Institute of Modern Physics, Chinese Academy of Sciences, Lanzhou 730000, China}
\address[d]{High Energy Accelerator Research Organization (KEK), Ibaraki 305-0801, Japan}
\address[xyz]{Department of Physics and Astronomy, Iowa State University, Ames, Iowa 50011, USA} 

\author[]{\\\vspace{0.2cm}(BLFQ Collaboration)}

\cortext[cor1]{Corresponding author}

\begin{abstract}
We solve for the $\rho$-meson's wave functions from a light-front QCD Hamiltonian determined for its constituent quark-antiquark and quark-antiquark-gluon Fock components, with a three-dimensional confinement using basis light-front quantization. From this, we obtain the leading-twist valence quark's parton distribution functions and transverse momentum-dependent parton distributions inside the $\rho$-meson. These results are qualitatively consistent with those of other models. We also demonstrate the important effects of a dynamical gluon on the $\rho$-meson's gluon densities, helicity, transversity, and tensor polarized distributions.
\end{abstract}
\end{frontmatter}

\section{Introduction}
The partonic structure of the $\rho$-meson plays an important role in our detailed understanding of hadronic physics since it is the third-lightest strongly interacting system following the pion and the kaon. In meson-exchange models, the $\rho$-meson plays a significant role in carrying the nuclear force within atomic nuclei. Despite having a similar quark content to the pion and being a spin-excited state of the latter, the $\rho$-meson is notably heavier. A key question arises: Why is this the case? 
The answer lies in the difference in strength of the color force among the partons inside each composite system.
%
In a broad sense, the mass and structural differences between the pion and the $\rho$-meson represent opportunities to probe the dynamical interplay of quark confinement, hyperfine splitting from gluon exchanges and chiral symmetry breaking all arising from QCD.

As a step in this direction, this work aims to anatomize the partonic structure of the $\rho$-meson in momentum space. This  can be attained through the study of the transverse momentum-dependent distribution functions (TMDs)~\cite{Collins:2011ca, Angeles-Martinez:2015sea, Diehl:2015uka} and the collinear parton distribution functions (PDFs)~\cite{Collins:1981uw, Gluck:1994uf, Martin:1998sq, Gluck:1998xa}.
TMDs are the three-dimensional (3-d) functions  that provide information about the longitudinal momentum fraction $(x)$ and transverse momentum $({\bf k}_\perp)$ carried by a parton inside a hadron. Such distributions are convoluted as nonperturbative quantities in the amplitude of the Drell-Yan (DY)~\cite{Ralston:1979ys, Donohue:1980tn, Collins:1984kg, Tangerman:1994eh, Zhou:2009jm, Collins:2002kn} and the semi-inclusive deep inelastic scattering (SIDIS) processes~\cite{Brodsky:2002cx, Ji:2004wu, Bacchetta:2017gcc}. The inadequacy of PDFs to provide the probability interpretation of partons in transverse momentum plane, makes them the limited versions of TMDs. These functions are, in principle, accessible through deep inelastic scattering (DIS) which is out of experimental reach for the $\rho$-meson due to its large width. Nevertheless, there is notable interest in investigating the momentum space distribution of the $\rho$-meson using various theoretical approaches~\cite{Ninomiya:2017ggn, Kaur:2020emh, Shi:2022erw}. Lower moments of the structure functions of $\rho$-meson have been explored using lattice QCD~\cite{Best:1997qp, Loffler:2021afv}.

Theoretical analyses of DIS processes, where the target or the produced hadron is a spin-1 hadron can be found in Refs.~\cite{Bacchetta:2000jk, Bacchetta:2001rb}. Hino and Kumano have provided the formalism to study the structure functions from polarized DY processes with a spin-1 hadron~\cite{Hino:1998ww, Hino:1999qi}. A total of nine valence quark TMDs and four PDFs, which are even in time-reversal (T-even), have been identified for spin-1 hadrons~\cite{Bacchetta:2001rb}. The special 1-d distribution function for the spin-1 composite systems, i.e. the tensor polarized function $(f_{1LL}(x))$, has drawn immense attention~\cite{Hoodbhoy:1988am, Close:1990zw, Umnikov:1996qv, Detmold:2005iz, Kumano:2010vz, Islam:2012zua, Miller:2013hla}. 
Additionally, four T-even gluon PDFs and thirteen gluon TMDs have been discovered for spin-1 systems at the leading-twist~\cite{Boer:2016xqr}. Positivity bounds on gluon TMDs for spin-1 hadrons have been discussed in Ref.~~\cite{Cotogno:2017puy}. 

Several theoretical approaches have aimed to depict the structure of the $\rho$-meson in momentum space, but very few are able to provide illumination from QCD first principles~\cite{Best:1997qp, Loffler:2021afv}. Additionally, viewing the hadron in terms of its partons from Euclidean methods is not as direct as that from the Minkowski space-time framework. 
Shi {\it et al.} have extracted the $\rho$-meson wave functions from the covariant Bethe-Salpeter Equation (BSE) and projected them over the light-front (LF) to generate the tomography of the $\rho$-meson~\cite{Shi:2022erw, Shi:2023oll}.

The Basis Light-Front Quantization (BLFQ) approach~\cite{Vary:2009gt} provides the hadronic light-front wave functions (LFWFs) from a Hamiltonian  derived from light-front QCD (LFQCD) and supplemented by a confinement term. These wave functions can be directly employed to investigate the partonic structure of hadrons. Notably, the BLFQ approach has been successfully applied to various QCD-bound state systems~\cite{Lan:2019vui, Lan:2019rba, Mondal:2019jdg, Xu:2021wwj, Mondal:2021czk, Lan:2021wok, Liu:2022fvl, Hu:2022ctr, Peng:2022lte, Xu:2022abw, Zhu:2023lst, Zhu:2023nhl, Kaur:2023lun, Lin:2023ezw}. 
Here, we extend our investigation to a higher-spin system, the $\rho$-meson, to compute its  quark and gluon distributions. 
We adopt an effective LF Hamiltonian and solve for its mass eigenstates at the scale suitable for low-resolution probes.  Our Hamiltonian incorporates light-front QCD interactions relevant to
constituent quark-antiquark and quark-antiquark-gluon Fock components of the mesons and a complementary 3-d confinement.  After fitting Hamiltonian parameters to mass spectra of unflavored light mesons~\cite{Lan:2021wok}, we compute the valence quark PDFs and TMDs, as well as the gluon PDFs of the  $\rho$-meson from the wave functions attained as eigenvectors of the Hamiltonian.

\section{$\rho$-meson LFWFs in the BLFQ framework}

To obtain the meson LFWFs in BLFQ, we solve the eigenvalue problem of the LF Hamiltonian: $P^+ P^- \ket{\Psi} = M^2 \ket{\Psi}$, where $P^+$ and $P^-$ represent the longitudinal momentum and LF Hamiltonian, respectively, acting upon the meson's state $\ket{\Psi}$. This state is expanded in the Fock space to include various components of quarks ($q$), anti-quarks ($\bar{q}$) and gluons ($g$)~\cite{Brodsky:1997de}, such that
\begin{equation}
\ket{\Psi} = \psi_{(q\bar{q})} \ket{q \bar{q}} + \psi_{(q\bar{q} g)} \ket{q \bar{q} g} + \psi_{(q\bar{q} q \bar{q})} \ket{q \bar{q} q \bar{q}} .\,.\,.\,,
\label{Eq:Fock-configurations}
\end{equation}  
with the probability amplitudes $\psi_{(...)}$ of various partonic configured states ($\ket{...}$) in the meson. Here, we retain the first two Fock components in Eq.~\eqref{Eq:Fock-configurations}, i.e., $\ket{q \bar{q}}$  and  $\ket{q \bar{q} g}$.

The LF Hamiltonian is considered as: $P^-=P^-_{\rm QCD} + P^-_{\rm C}$, where the LFQCD Hamiltonian with one dynamical gluon in the LF gauge ($A^+=0$) is given by~\cite{Lan:2021wok, Brodsky:1997de}
\begin{align}
P^-_{\rm QCD} =&\int \frac{{\rm d}^2{x^\perp} {\rm d}x^-}{2} \bigg[\bar{\psi}\gamma^+ \frac{m_0^2 +(\iota \partial^\perp)^2}{\iota \partial^+} \psi \nonumber\\ +& A_a^i \left(m_g^2 + (\iota \partial^\perp)^2 \right)  A_a^i + 2 g_s \bar{\psi} \gamma_\mu T^a A_a^\mu \psi \nonumber\\
+& g_s^2 \bar{\psi} \gamma^+ T^a \psi \frac{1}{(\iota \partial^+)^2} \bar{\psi} \gamma^+ T^a \psi \bigg]\;.
\label{Eq:QCD-Hamiltonian}
\end{align}
The first two terms in above equation are the kinetic energies with $m_0$ and $m_g$ being the bare masses of quark and gluon, respectively. $\psi$ and $A^\mu$ denote the quark and gluon fields, respectively. The variables $x^-$ and $x^\perp$ are the longitudinal
and transverse position coordinates, respectively. The interaction between the partons inside the meson is incorporated in the remaining terms with $g_s$ being the coupling constant. $T^a$ represents the adjoint matrices of SU(3) gauge group, and $\gamma^\mu$ denotes the Dirac matrices. For the quark mass correction due to quantum fluctuations to a higher Fock sector, we introduce a mass counter term,
$\delta m_q = m_0 - m_q$, for the quark in the leading Fock component with $m_q$ being the renormalized quark mass~\cite{Glazek:1993rc, Karmanov:2008br, Karmanov:2012aj, Zhao:2014hpa, Zhao:2020kuf}. Apart from this, a mass parameter $m_f$ is introduced to parameterize the nonperturbative effects in the vertex interactions~\cite{Glazek:1992aq, Burkardt:1998dd}.

We employ the following confinement in the leading Fock sector, which includes transverse and longitudinal confining potentials~\cite{Li:2015zda, Lan:2021wok, Zhu:2023lst, Xu:2023nqv}
\
\begin{equation}
P_{\rm C}^- P^+ = \kappa^4 \left( x(1-x) {\bf r}^2_\perp -\frac{\partial_x(x(1-x)\partial_x)}{(m_q+m_{\bar{q}})^2} \right)\;,
\end{equation}      
with $\kappa$ being the confining strength and ${\bf r}_\perp$ is the holographic variable~\cite{Brodsky:2014yha}, which measures the transverse separation between the valence quarks~\cite{Brodsky:2014yha}. Confinement
in the $|q\bar{q}g\rangle$ sector is implemented through the massive gluon (compatible with the low mass scale of the model) and the limited basis in the transverse direction discussed below~\cite{Lan:2021wok, Zhu:2023lst,Xu:2023nqv}.

For the basis of BLFQ, a discretized plane-wave basis and a 2-d harmonic oscillator (HO) wave functions are employed to describe the longitudinal and transverse dynamics of Fock particles, respectively. The longitudinal motion is confined to a 1-d box of length $2L$ with antiperiodic (periodic) boundary conditions for fermion (boson). The longitudinal momentum of the particle is then defined as $p^+=2\pi k / L$, where $k$ takes half-integer (integer) values for a fermion (boson). We neglect the zero mode for the boson.  For the many-body basis states, the total longitudinal momentum becomes $P^+=\sum_i p_i^+=2 \pi K/L$, where $K= \sum_i k_i$. The longitudinal momentum fraction of the $i$th parton is expressed as $x=p_i^+/P^+ = k_i/K$. The 2-d HO basis function $\phi_{n,m}({\bf k}_\perp; b)$, with scale parameter $b$, takes different values of the radial and angular momentum quantum numbers denoted by $n$ and $m$. 
Therefore, a single parton's basis state is expressed by the set of four quantum numbers, $\bar{\alpha}= \left\lbrace x, n, m, \lambda \right\rbrace$, where
the parton's helicity is denoted by $\lambda$. The many-body basis states are identified as the direct product of the single particle basis states $\ket{\alpha}=\otimes \ket{\alpha_i}$. Further, for many-body basis states, the total angular momentum projection is defined as $\Lambda= \sum_i(m_i+\lambda_i)$.

We introduce two truncation parameters in the transverse and longitudinal directions, $N_{\rm max}$ and $K$, respectively to truncate the infinite basis of each Fock sector. The transverse truncation is set by $N_{\rm max} \geq \sum_i (2 n_i + | m_i | + 1)$~\cite{Zhao:2014xaa}. The parameter $K$ is the resolution in the longitudinal direction. $N_{\rm max}$ regulates the ultra-violet (UV) and infrared (IR) regions by introducing the cutoffs, $\Lambda_{\rm UV}=b \sqrt{N_{\rm max}}$ and $\Lambda_{\rm IR}=b /\sqrt{N_{\rm max}}$, respectively~\cite{Zhao:2014xaa}.

The resulting LFWFs in momentum space are then expressed as 
\begin{equation}\label{WF}
\Psi^\Lambda_{\lbrace{\mathcal{N};\lambda_i\rbrace}}(x,{\bf k}_{\perp i}) \propto \sum_{n_i,m_i} \psi^{\mathcal{N}}(\left \lbrace \bar{\alpha_i} \right \rbrace) \prod_{i=1}^{\mathcal{N}} \phi_{n_i,m_i}({\bf k}_{\perp i}; b)\;,
\end{equation} 
where $\psi^{\mathcal{N}=2}$ and $\psi^{\mathcal{N}=3}$ are the components of the eigen solutions  associated with the Fock sectors $\ket{q \bar{q}}$ and $\ket{q \bar{q} g}$, respectively, obtained after diagonalizing the full Hamiltonian matrix in the BLFQ
framework.

\section{Quark TMDs of the $\rho$-meson}
The quark TMDs are parameterized through the quark-quark correlation function defined as~\cite{Tangerman:1994eh, Ninomiya:2017ggn, Bacchetta:2000jk, Mulders2001TransverseMD, Bacchetta:2001rb}
\begin{equation}
\Phi^{q[\Gamma]} (x,{\bf k}_\perp) = \frac{1}{2}{\rm Tr}\left(\Gamma \Phi^{q(\Lambda)_S}(x,{\bf k}_\perp) \right)\;,
\label{Eq:Relation-Dirac}
\end{equation}
where the Dirac matrix $\Gamma$ determines the Lorentz structure of the correlator and its twist, and the two-quark correlator function $\Phi^{q(\Lambda)_S}_{i j}(x,{\bf k}_\perp)$ is defined as
\begin{align}
& \Phi^{q(\Lambda)_S}_{i j}(x,{\bf k}_\perp) =
 \int \frac{{\rm d}z^- \, {\rm d}^2 {\bf z}_\perp}{(2\pi)^3} \, 
e^{\iota k \cdot z} \nonumber\\
& \times {}_S\langle P,\Lambda | \bar{\psi}_{j}(0) U_{[0,z]}
\psi_{i}(z^-,{\bf z}_\perp) | P, \Lambda \rangle_S \big{\vert}_{z^+=0}\;.
\label{Eq::correlator}
\end{align}
The gauge link $U_{[0,z]}$ in the above Eq.~\eqref{Eq::correlator} ensures the gauge invariance of the bilocal quark field operators in the correlation function. Note that this work is limited to study the T-even TMDs at the leading-twist, which in practice, can be extracted by taking the gauge link to be unity~\cite{Pasquini:2014ppa, Noguera:2015iia, Ahmady:2019yvo, Kaur:2020emh, Shi:2022erw}. $P$ is the momentum of the target, and 
 $\Lambda=0,\pm1$ represents the spin projection of the target in the longitudinal direction.

With different Dirac matrices, $\Gamma=\gamma^+$, $\gamma^+\gamma_5$, $\iota \sigma^{i+}\gamma^5$, one  can parameterize the distribution functions at the leading-twist~\cite{Bacchetta:2001rb, Ninomiya:2017ggn},

\begin{align}
&\Phi^{q[\gamma^+]}(x,{\bf k}_\perp)
 = 
f^q_1(x, {k}_\perp^2) + S_{LL} \,
f^q_{1LL}(x, {k}_\perp^2) \nonumber\\
&+ \frac{\boldsymbol{S}_{LT} \cdot {\bf k}_\perp}{M}\, f^q_{1LT}(x, {k}_\perp^2) +\frac{{\bf k}_\perp \cdot \boldsymbol{S}_{TT} \cdot {\bf k}_\perp}{M^2} \, f^q_{1TT}(x, {k}_\perp^2)\, ,  \label{Eq:unpol-quark}\\
&\Phi^{q[\gamma^+\gamma_5]}(x,{\bf k}_\perp)  = S_L\,g^q_{1L}(x,{k}_\perp^2)+ \frac{{\bf k}_\perp \cdot \boldsymbol{S}_\perp}{M} g^q_{1T}(x,{k}_\perp^2)\, , \label{Eq:longi-quark}\\
&\Phi^{q[\iota \sigma^{i+}\gamma_5]} (x,{\bf k}_\perp)  = S_\perp^i h^q_1(x,{k}_\perp^2) 
+ S_L\frac{k_\perp^i}{M}h_{1L}^{\perp q}(x,{k}_\perp^2) \nonumber\\
&+
\frac{1}{2\,M^2}
\left(2\,k_\perp^i {\bf k}_\perp \cdot \boldsymbol{S}_\perp - S_\perp^i~{k}_\perp^2\right) h_{1T}^{\perp q}(x,{k}_\perp^2)\,,
\label{Eq:trans-quark}
\end{align}

with 
 \begin{align}
S_{LL}&=\left(3 \Lambda^2-2 \right)\left(\frac{1}{6}-\frac{1}{2} S_L^2\right),\\
S^i_{LT}&=\left(3 \Lambda^2-2 \right)S_L S^i_\perp,\\
S_{TT}^{ij}&=\left(3 \Lambda^2-2 \right)(S_\perp^i S_\perp^j-\frac{1}{2}S_\perp^2~ \delta^{ij})\;.
\label{Eq:spins}
\end{align}
Here, $S_L$ and $S^{i(j)}_\perp$ represent the longitudinal and transverse polarization of the target, respectively, where $i(j)$ can take the  direction along either $x$ or $y$ axis. 

\subsubsection*{Overlap description of the TMDs}
Substituting the meson state and the quark field operator in Eq.~\eqref{Eq::correlator}, and then using Eqs.~\eqref{Eq:unpol-quark}-\eqref{Eq:trans-quark}, gives rise to the TMDs in terms of overlap of the LFWFs. We obtain
%
 \begin{align}
  f^q_1&= \frac{1}{3} \sum_{\mathcal{N}=2}^{3} \left[ A^{++}_{\mathcal{N};++} + A^{00}_{\mathcal{N};++} + A^{--}_{\mathcal{N};++}\right]\;, \label{eq:f_1} 
 \\
  g^q_{1L}&= \frac{1}{2} \sum_{\mathcal{N}=2}^{3} \left[ A^{++}_{\mathcal{N};++} - A^{++}_{\mathcal{N};--} \right]\;,
\\
  h_1^q &= \frac{1}{\sqrt{2}} \sum_{\mathcal{N}=2}^{3} A^{+0}_{\mathcal{N};+-} \;,
  \\
  g^q_{1T}&= \frac{M}{\sqrt{2} k_\perp^2} \sum_{\mathcal{N}=2}^{3} \Re \left[k_R \left( A^{+0}_{\mathcal{N};++} + A^{0-}_{\mathcal{N};++} \right) \right] \;,
\\
  h_{1L}^{\perp q} &= -\frac{M}{k^2_\perp} \sum_{\mathcal{N}=2}^{3} \Re \left[k_R ~ A^{--}_{\mathcal{N};-+}  \right] \;,\\
  h_{1T}^{\perp q} &= \frac{\sqrt{2}M^2}{k_\perp^4} \sum_{\mathcal{N}=2}^{3} \Re \left[k^2_R ~ A^{0-}_{\mathcal{N};-+}  \right] \;,
\\
  f^q_{1LL} &= \sum_{\mathcal{N}=2}^{3} \left[A^{00}_{\mathcal{N};++} - \frac{1}{2} \left( A^{++}_{\mathcal{N};++} + A^{--}_{\mathcal{N};++} \right)\right]\;,  
 \\
  f^q_{1LT} &= \frac{M}{\sqrt{2} k_\perp^2} \sum_{\mathcal{N}=2}^{3} \Re \left[k_R \left( A^{+0}_{\mathcal{N};++} - A^{0-}_{\mathcal{N};++} \right) \right] \;, \\
  f^q_{1TT} &= \frac{M}{k^4_\perp} \sum_{\mathcal{N}=2}^{3} \Re \left[ k_R^2 ~ A^{+-}_{\mathcal{N}; ++} \right]\;, \label{eq:f_1LT}
 \end{align}

 where 
  \begin{widetext}
 \begin{align}
 A^{\Lambda^\prime, \Lambda}_{2;\lambda^\prime_q\lambda_q} &= \sum_{\lambda_{\bar{q}}} \int\frac{\prod_{i=1}^2 {\rm d}x^\prime_i {\rm d}^2{\bf k}^\prime_{\perp i}}{[2(2\pi)^3]^2}2(2\pi)^3 ~ \Psi^{*\Lambda^\prime}_{2;\lambda^\prime_q \lambda_{\bar{q}}} \Psi^{\Lambda}_{2;\lambda_q \lambda_{\bar{q}}} \delta^3 \left(\tilde{k}^\prime_1 - \tilde{k}\right) \delta^3 \left(\tilde{P} -\sum_i \tilde{k}_i^\prime \right)\;, \label{eq:overlap-qqbar}\\
 A^{\Lambda^\prime, \Lambda}_{3;\lambda^\prime_q\lambda_q} &= \sum_{\lambda_{\bar{q}}, \lambda_g} \int\frac{\prod_{i=1}^3 {\rm d}x_i {\rm d}^2{\bf k}_{\perp i}}{[2(2\pi)^3]^3}2(2\pi)^3  ~ \Psi^{*\Lambda^\prime}_{3;\lambda^\prime_q \lambda_{\bar{q}}\lambda_g} \Psi^{\Lambda}_{3;\lambda_q \lambda_{\bar{q}}\lambda_g} \delta^3 \left(\tilde{k}_1 - \tilde{k}\right) \delta^3 \left(\tilde{P} -\sum_i \tilde{k}_i \right)\;, \label{eq:overlap-qqbarg}
 \end{align}
  with $\delta^3\left(\tilde{P} -\sum_i \tilde{k}_i \right)$ = $\delta\left(1-\sum_i x_i \right) \delta^2 \left({\bf P}_\perp - \sum{\bf k}_{\perp i} \right)$. 
\end{widetext}
We omit the arguments ($\tilde{k}^\prime_1,\,\tilde{k}^\prime_2$) from the quark-antiquark LFWFs
and ($\tilde{k}_1,\,\tilde{k}_2,\, \tilde{k}_g$) from the quark-antiquark-gluon LFWFs, where $\tilde{k}_i\equiv (x_i,\,{\bf k}_{\perp i})$. Note that the TMDs are defined in Eqs.~\eqref{eq:f_1}-\eqref{eq:f_1LT} in terms of the helicity amplitudes denoted by $A^{\Lambda^\prime \Lambda}_{\mathcal{N};\lambda_q^\prime \lambda_q}$.
The overlap representation of the T-even $\rho$-meson TMDs in terms of the helicity amplitudes has also been defined in Refs.~\cite{Kaur:2020emh, Shi:2022erw}. However, those definitions do not involve the contribution from the higher Fock sector $\ket{q \bar{q} g}$. 

\begin{figure*}
\includegraphics[scale=0.58]{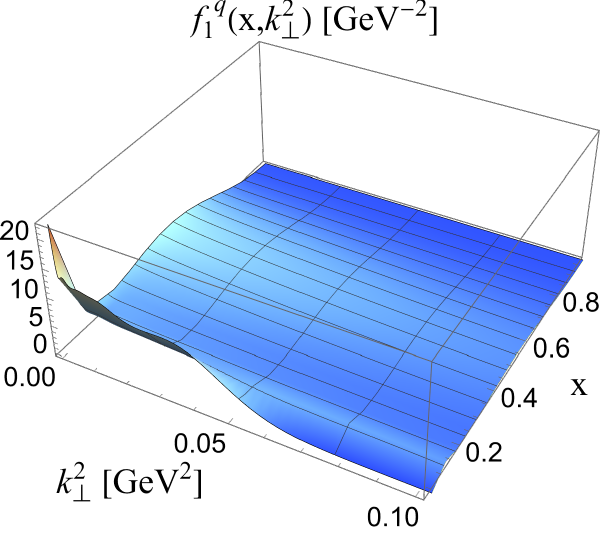} \hspace{0.1cm}
\includegraphics[scale=0.58]{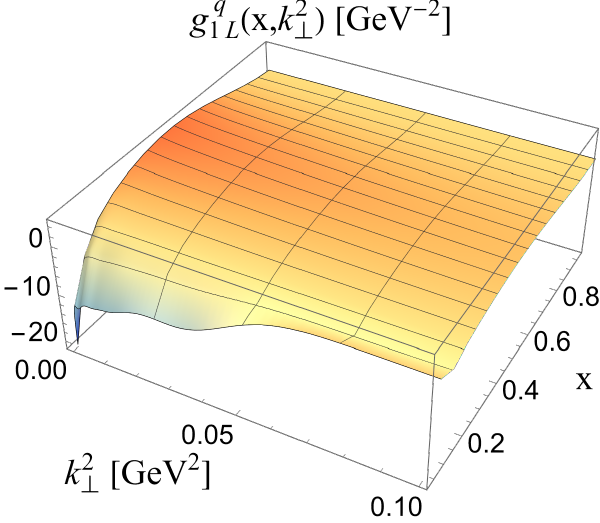} \hspace{0.1cm}
\includegraphics[scale=0.58]{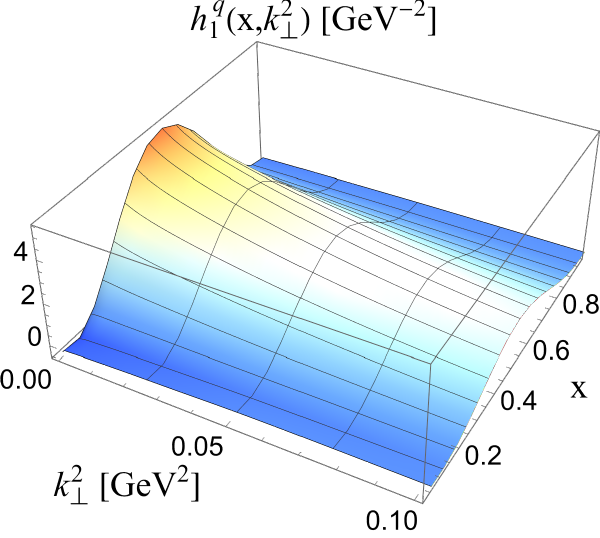} \\ \\
\includegraphics[scale=0.58]{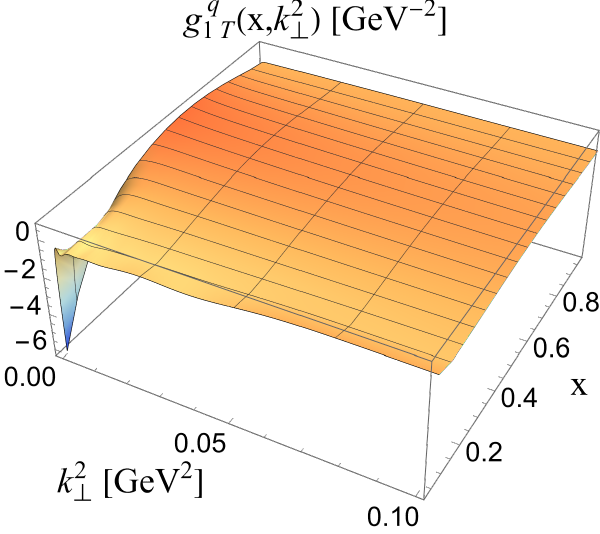}\hspace{0.1cm}
\includegraphics[scale=0.58]{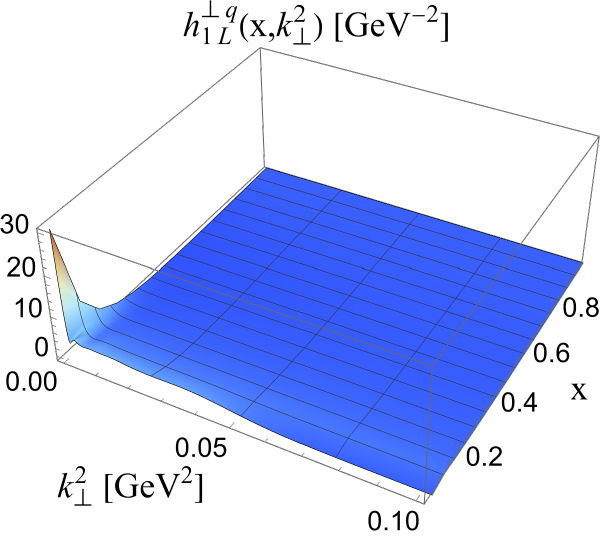} \hspace{0.1cm}
\includegraphics[scale=0.58]{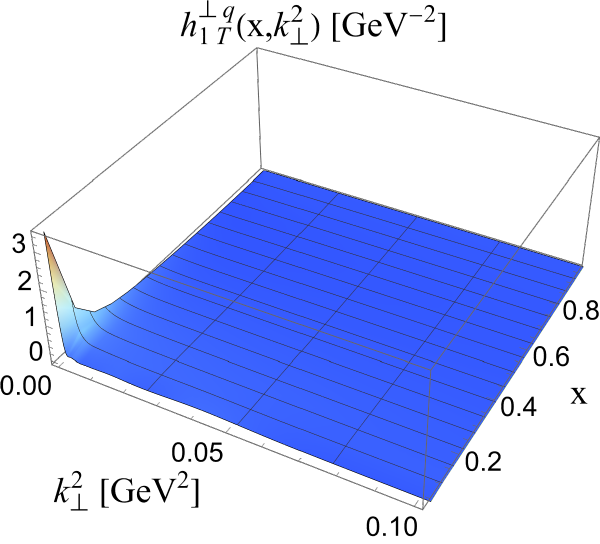}\\ \\
\includegraphics[scale=0.58]{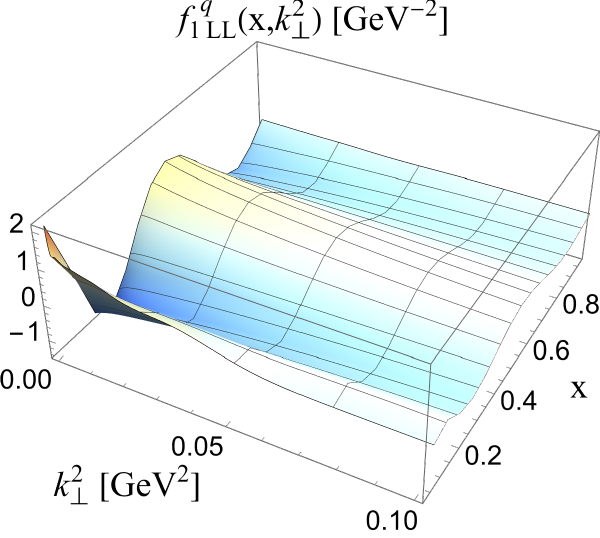}\hspace{0.1cm}
\includegraphics[scale=0.58]{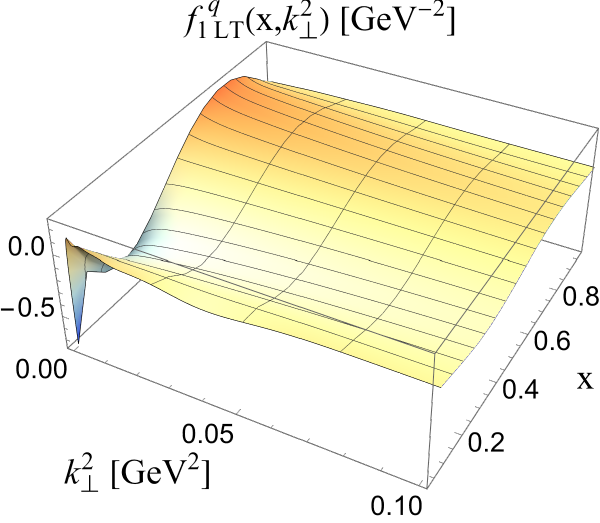}\hspace{0.1cm}
\includegraphics[scale=0.58]{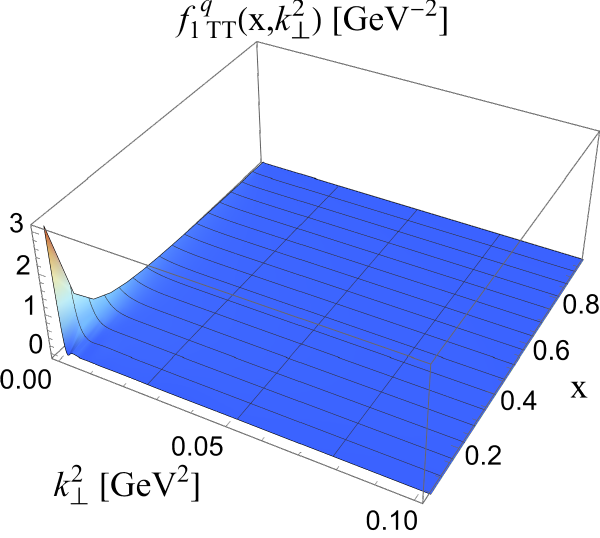}
\caption{The valence quark TMDs for the $\rho$-meson as a function of $x$ and $k^2_\perp$.}
\label{fig:3D-TMDs}
\end{figure*}

For the numerical calculations, we fix the parameters, as summarized in Table~\ref{table:parameters}, by fitting the mass spectra of unflavored light mesons~\cite{Lan:2021wok, Zhu:2023lst}. Besides this, the truncation parameters are chosen as $\lbrace N_{\rm max}, K\rbrace=\lbrace 14, 15\rbrace$. The resulting wave functions of the $\rho$-meson have three components: $S$, $P$ and $D$ waves, which correspond to the relative orbital angular momentum (OAM)  $L_z=0,\,\pm 1$, and $\pm 2$, respectively, where $L_z=\Lambda-(\lambda_q+\lambda_{\bar{q}})$ for the $|q\bar{q}\rangle$ sector and $L_z=\Lambda-(\lambda_q+\lambda_{\bar{q}}+\lambda_g)$ for the $|q\bar{q} g\rangle$ sector. The contributions of these components in the LFWFs with different $\Lambda\,(=0,\pm 1)$ for the Fock sectors under consideration ($\ket{q \bar{q}}$ and $\ket{q \bar{q} g}$) is listed in Table~\ref{table:wave-contribution}. We observe that the probability of the $\ket{q \bar{q} g}$ Fock sector is slightly larger than that of the $\ket{q \bar{q}}$ Fock sector.
 \begin{table}[hbt!]
 \footnotesize
 \caption{The parameters used in this work.}
\label{table:parameters}
\vspace{0.2cm}
 \centering
 \begin{tabular}{ c  c   c   c   c   c }
 \hline\hline
  $m_q$ & $m_g$ & $\kappa$ & $m_f$ & $b$ & $g_s$  \\
 $[{\rm GeV}]$ & $[{\rm GeV}]$ & $[{\rm GeV}]$ & $[{\rm GeV}]$ & $[{\rm GeV}]$ & \\
 \cline{1-6} 
 0.39 & 0.60 & 0.65 & 5.69 & 0.29 & 1.92 \\ 
 \hline \hline
\end{tabular}

\end{table}

\begin{table}[hbt!]
\footnotesize
\caption{The probabilities of different wave components in the LFWFs for different Fock sectors.}
\label{table:wave-contribution}
\vspace{0.2cm}
\centering
\begin{tabular}{c | l l | l l}
\hline \hline
\multirow{2}{*}{Wave component} &
\multicolumn{2}{c}{$\ket{q \bar{q}}$} &
\multicolumn{2}{c}{$\ket{q \bar{q} g}$} \\ \cline{2-5}
&  $\Lambda=0$ & $\Lambda=1$ &  $\Lambda=0$ & $\Lambda=1$ \\
\hline
$S$ & $48.6\%$ & $49.14\%$ & $50.56\%$ & $50.12\%$ \\
$P$ & $0.008\%$ & $0.039\%$ & $0.83\%$ & $0.69\%$ \\
$D$ & $-$ & $\approx 10^{-6}\%$ & $\approx 10^{-5}\%$ & $\approx 10^{-4}\%$ \\
\hline \hline
\end{tabular}
\end{table}


In Fig.~\ref{fig:3D-TMDs}, we show the valence quark TMDs of $\rho$-meson at the model scale $\mu^2_0 = 0.34 $ GeV$^2$ with respect to the squared of the transverse momentum ${k}_\perp^2$ and the longitudinal momentum fraction $x$ carried by the active quark. In the upper panel of Fig.~\ref{fig:3D-TMDs}, we present the unpolarized, helicity-dependent and transversity distributions, which are diagonal in OAM in their overlap representation of the LFWFs. These TMDs have their PDF limits. We observe that the TMDs $f_1$ and $g_{1L}$ exhibit their peaks at small-$x$, indicating the effects from incorporating a dynamical gluon. It seems to begin to account for the effects of evolution as one may expect. 
Meanwhile, for the TMD $h_1$, the absence of a peak at low-$x$  is attributed to the fact that it involves
the overlap of the LFWFs corresponding to  $\Lambda=1$ and $\Lambda=0$.
The TMDs obtained in our BLFQ approach are qualitatively consistent with other model predictions~\cite{Ninomiya:2017ggn, Kaur:2020emh, Shi:2022erw}, except in the low-$x$ region. The potential reason for this difference at small-$x$ region is the inclusion of the $|q\bar{q}g\rangle$ Fock sector in our approach,
while the TMDs computed in other models are based on the leading Fock component~\cite{Ninomiya:2017ggn, Kaur:2020emh, Shi:2022erw}. 

The TMD $g_{1T}~(h^\perp_{1L})$, which describes the probability of finding a longitudinally (transversely) polarized quark in a transversely (longitudinally) polarized $\rho$-meson, is shown in the middle panel of Fig.~\ref{fig:3D-TMDs}. Again, our distributions find qualitative agreement with other model predictions at $x>0.2$ for small ${k}^2_\perp$, where $g_{1T}$ is positively distributed and $h^\perp_{1L}$ is negative~\cite{Ninomiya:2017ggn, Kaur:2020emh, Shi:2022erw}. 
Additionally, we observe a non-vanishing $h_{1T}^\perp$ similar to that obtained in the light-front quark model (LFQM)~\cite{Kaur:2020emh} and BSE approach~\cite{Shi:2022erw}. However, our distribution is comparatively smaller in magnitude. This difference is attributed to the superposition of the $S$-wave  on $P$ and $D$-wave components for $h^\perp_{1L}$ and $h^\perp_{1T}$, respectively. Note that the $P$ and $D$-wave components of the LFWFs are found to be very weak in our approach, as illustrated in Table \ref{table:wave-contribution}. 

In the lower panel of Fig~\ref{fig:3D-TMDs}, we present the tensor polarized $\rho$-meson TMDs with the valence quark being unpolarized: $f_{1LL}, f_{1LT}$, and $f_{1TT}$. Similar to other TMDs, we observe a peak in all the tensor polarized TMDs in the small-$x$ region, arising from the contribution of the Fock sector $\ket{q\bar{q}g}$. At $x=0.5$, the distribution $f_{1LL}$ exhibits maxima when $k^2_\perp\approx 0$, and $f_{1LT}$ is zero, being independent of the quark's transverse momentum. Such features of these special TMDs have also been observed in other models~\cite{Ninomiya:2017ggn, Kaur:2020emh}. The explicit inclusion of the Fock sector beyond the valence sector has made these TMDs asymmetric in $x$. Additionally, $f_{1TT}$ remains non-zero only when the valence quark carries small momentum in the transverse direction. However, other model predictions have depicted the quark's distribution over comparatively larger values of ${k}^2_\perp$~\cite{Ninomiya:2017ggn, Kaur:2020emh, Shi:2022erw}. Note that our  valence quark TMDs satisfy all the positivity constraints discussed in Ref.~\cite{Bacchetta:2001rb}.

\section{PDFs of the $\rho$-meson}
PDFs, as 1-d distribution functions, provide a probabilistic interpretation of the parton carrying the longitudinal momentum fraction. In practice, these functions can be retrieved by integrating out the parton's transverse momentum from the TMDs~\cite{Ninomiya:2017ggn, Kaur:2020emh, Shi:2022erw}.
Here, we present the valence quark and gluon PDFs within the $\rho$-meson.    
\subsection{Valence quark PDFs}
The quark-quark correlator for PDFs is defined as~\cite{Bacchetta:2000jk}
\begin{align}
& \Phi^{q(\Lambda)_S}_{ij}(x) =  \int \frac{{\rm d}z^-}{2\pi} \, 
e^{\iota k \cdot z} \nonumber\\
& \times {}_S\langle P,\Lambda | \bar{\phi}_{j}(0) U_{[0,z]}
\phi_{i}(z^-) | P, \Lambda \rangle_S \big{\vert}_{\substack{z^+=0 \\ {\bf z}_\perp=0}}\;.
\end{align}
With the LF gauge, the gauge link is unity. The projection of different Dirac matrices parameterizes the PDFs as~\cite{Bacchetta:2000jk, Hino:1999qi}
\begin{align}
\Phi^{q[\gamma^+]}(x) &= f_1^q(x) +S_{LL} f_{1LL}^q(x)\;,\\
\Phi^{q[\gamma^+\gamma_5]}(x) &= S_L g_{1L}^q(x)\;, \\
\Phi^{q[\iota \sigma^{i+}\gamma_5]}(x) &= S_\perp^i h_1^q(x)\;.
\end{align}

\begin{figure}
\centering
\includegraphics[scale=0.38]{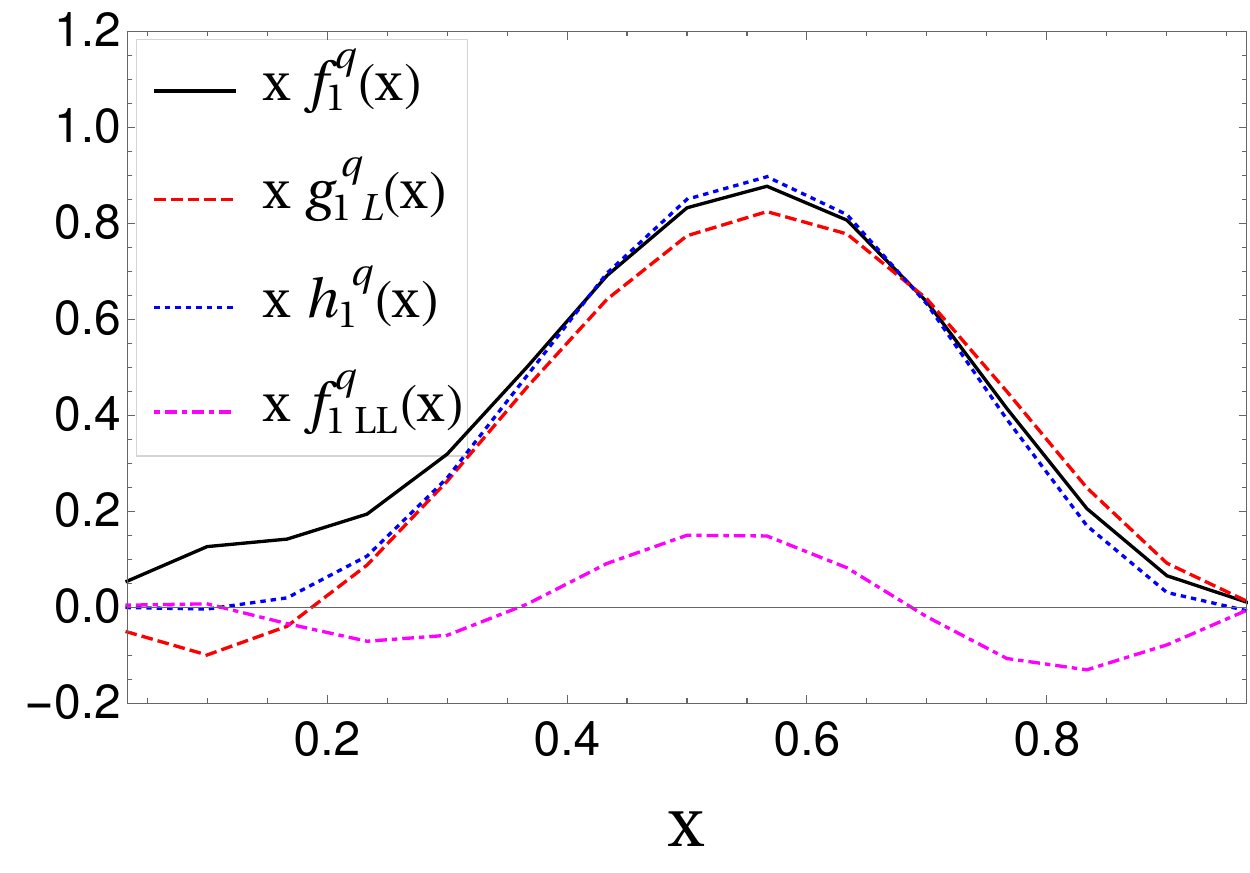}
\caption{The valence quark PDFs as a function of $x$ evaluated at the model scale.}
\label{fig:quark-PDFs-mu0}
\end{figure} 

In Fig.~\ref{fig:quark-PDFs-mu0}, we show the valence quark PDFs of the $\rho$-meson $f_1, g_{1L}, h_1$ and $f_{1LL}$ at the model scale $\mu_0$. We observe a qualitative difference of the unpolarized PDF $f_1$ and helicity PDF $g_{1L}$ at the small-$x$ region compared to other theoretical approaches~\cite{Ninomiya:2017ggn, Kaur:2020emh, Shi:2022erw}. This is because of our adoption of the phenomenological mass parameter $m_f$ in the vertex interactions that imitates the nonperturbative effects from the higher Fock sectors~\cite{Burkardt:1998dd}. The qualitative behavior of our $h_1$ aligns well with other predictions in the literature~\cite{Ninomiya:2017ggn, Kaur:2020emh, Shi:2022erw}. 

The tensor structure of $\rho$-meson described by $f_{1LL}$ has shown significant variations among different models~\cite{Sun:2017gtz, Kaur:2020emh, Ninomiya:2017ggn, Mankiewicz:1988dk, Shi:2022erw}. The qualitative behavior of the $f_{1LL}$ in our BLFQ approach bears similarities to some recent models predictions~\cite{Ninomiya:2017ggn, Kaur:2020emh}. 

It is important to note that, at the model scale, our  PDFs satisfy the following sum rules
\begin{equation}
\int {\rm d}x f^q_1(x) = 1 \hspace{0.4cm}; \hspace{0.4 cm} \int {\rm d}x f^q_{1LL}(x)  = 0\;.
\end{equation}

At our model scale, we obtain the contribution from the valence quark to the spin and tensor charge of the $\rho$-meson as $\int dx g^q_{1L}(x) \approx 0.54$ and $\int dx h^q_1(x) \approx 0.70$, respectively. The remaining portion of these quantities is expected to arise from other partons inside the meson and from the OAM in the case of spin. Our predictions concerning the gluon contribution to these quantities are specified in Section-\ref{sec:g-PDFs}.

\subsection{Gluon PDFs}
\label{sec:g-PDFs}

\begin{figure}
\centering
\includegraphics[scale=0.36]{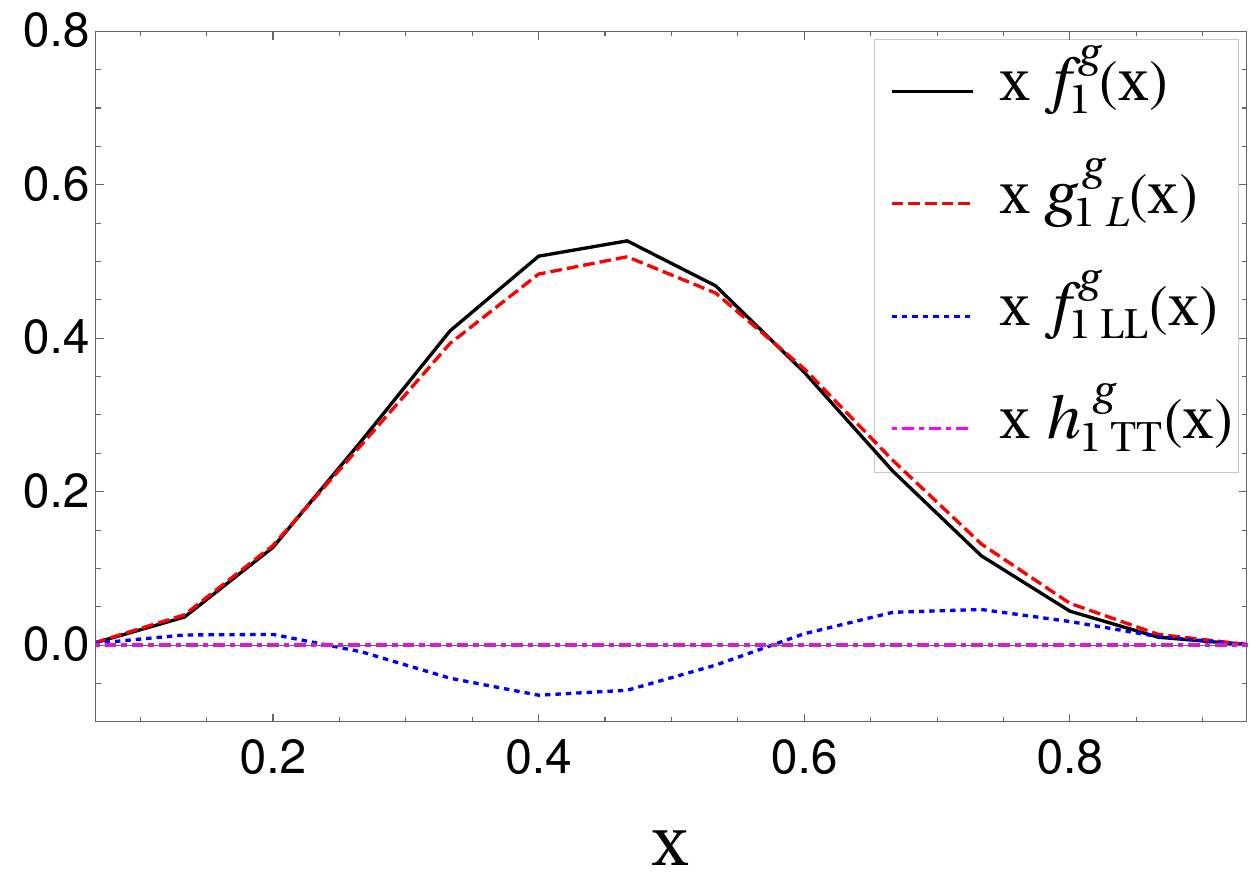}
\caption{The gluon PDFs as a function of $x$ evaluated at the model scale. The $h_{1TT}$ is near zero on this scale.}
\label{fig:gluon-PDFs-mu0}
\end{figure}

The gluon-gluon correlation for PDF in a  hadron is given by~\cite{Mulders2001TransverseMD}  
\begin{align}
&\Phi^{\mu \nu; \rho \sigma}(x) = \frac{1}{x P^+} \int \frac{{\rm d}z^-}{2\pi} \, 
e^{\iota k \cdot z} \nonumber\\
& \times {}_S\langle P,\Lambda | F^{\mu\nu}(0) U_{[0,z]} 
F^{\rho \sigma}U^\prime_{[0,z]} | P, \Lambda \rangle_S \big{\vert}_{\substack{z^+=0 \\ {\bf z}_\perp=0}}\;.
\label{eq:gluon-gluon-correlator}
\end{align}
The T-even gluon PDFs at the leading-twist are parameterized through the correlator, Eq.~\eqref{eq:gluon-gluon-correlator}, as~\cite{Cotogno:2017puy, Boer:2016xqr}
\begin{align}
\delta^{ij}_T\Phi^{g[ij]}(x) &= f_1^g(x) +  S_{LL} f^g_{1LL}(x)\;,\\
\iota \epsilon^{ij}_T \Phi^{g[ij]}(x) &= S_L g^g_{1L}(x)\;,\\
-\hat{S}\Phi^{g[ij]}(x) &= S^{ij}_{TT} h^g_{1TT}(x)\;,
\end{align}
with $\hat{S}$ being the symmetrization operator in $i$ and $j$~\cite{Meissner:2007rx}.

Fig.~\ref{fig:gluon-PDFs-mu0} shows the T-even gluon PDFs for the $\rho$-meson evaluated at the model scale. When compared with the quark PDFs, the peaks of the gluon unpolarized and helicity distributions shift towards the lower $x$, while the magnitude decreases. 
The tensor-polarized gluon PDF, denoted as $f^g_{1LL}$, exhibits an opposite trend  compared to that of the quark. Notably, the gluon transversity PDF $h_{1TT} (\propto\mathcal{A}^{-+}_{3;+-}$, where $\mathcal{A}^{\Lambda^\prime \Lambda}_{3;\lambda^\prime_g\lambda_g}$ defines the helicity amplitude analogous to Eq.~\eqref{eq:overlap-qqbarg}), which can only be seen for the spin $\geq 1$ systems~\cite{Jaffe:1989xy, Mulders:2000sh, Kumano:2022cdj}, vanishes in our current treatment of the BLFQ approach for the $\rho$-meson.  This is simply because of the very weak $D$-wave component in our wave functions as shown in Table-\ref{table:wave-contribution}. However, we anticipate that the $D$-wave component will become more apparent and strengthen as we include Fock sectors beyond $\ket{q \bar{q} g}$. Hence, this PDF would possibly survive. 

Additionally, it is noteworthy that approximately $50.2\%$ of the $\rho$-meson's spin arises from the gluon's helicity. Consequently, from the spin sum rule, it appears that the negative, though significantly less contribution of $\sim 4\%$ emerges from the OAM. 
 This proportion is uncertain, and the system may receive more contribution from the OAM as we include more gluons and sea quarks as well as their interactions. 
%
Our gluon PDFs follow the positivity constraints defined in Ref.~\cite{Cotogno:2017puy}. 

\subsection{QCD evolution}

\begin{figure}
\centering
\includegraphics[scale=0.38]{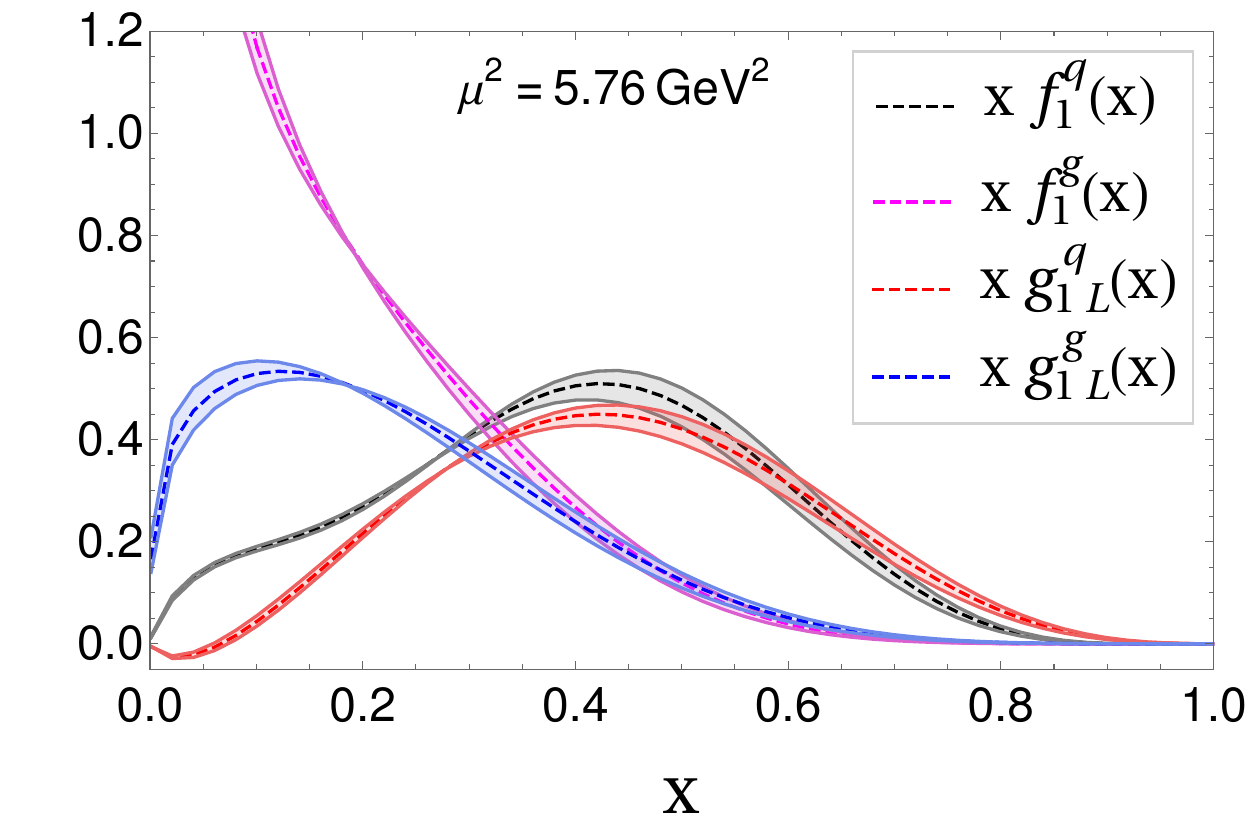}
\caption{The quark and gluon PDFs $f_1$ and $g_{1L}$ at scale $\mu^2=5.76$ GeV$^2$. 
The error bands in our evolved PDFs are due to the spread in the initial scale $\mu^2_0=0.34 \pm 0.034$ GeV$^2$.}
\label{fig:PDFs-mu}
\end{figure}

We perform QCD evolution using next-to-next-to-leading order (NNLO) Dokshitzer-Gribov-Lipatov-Altarelli-Parisi (DGLAP) equations to obtain the PDFs at a large scale, $\mu^2$~\cite{Dokshitzer:1977sg, Gribov:1972ri, Altarelli:1977zs}. 
Since the unflavored light mesons are the eigenstates of the same LF Hamiltonian Eq.~\eqref{Eq:QCD-Hamiltonian}, we assume our model scale for the $\rho$-meson to be same as that of the pion,  $\mu^2_0=0.34$ GeV$^2$~\cite{Lan:2021wok, Zhu:2023lst}. We consider $10\%$ uncertainty in the initial scale. The quark and gluon PDFs $f^{q/g}_1$ and $g^{q/g}_{1L}$ at $\mu^2=5.76$ GeV$^2$ are shown in Fig.~\ref{fig:PDFs-mu}. 

Further, the moments of the PDFs are defined as
\begin{equation}
\langle x^n \rangle_{{\rm \lbrace PDF \rbrace}^v} = \int dx ~ x^{n} {\lbrace {\rm PDF} \rbrace^v }\;,
\end{equation}
where $n=0,1,2,3,...$. We find that at $\mu^2=5.76$ GeV$^2$, $2 \braket{x}_{q}=0.47$, i.e., $47\%$ of the total momentum of the $\rho$-meson is carried by the valence quarks, as shown in Table~\ref{tab:moments-f1}. 
We compare the first three moments of the unpolarized and helicity PDFs for both the valence quarks and gluon with lattice QCD~\cite{Best:1997qp} and BSE approach~\cite{Shi:2022erw} in Tables~\ref{tab:moments-f1} and \ref{tab:moments-g1L}, respectively. Our predictions for $\langle x \rangle_{f_1/g_{1L}}$ are somewhat smaller in comparison with other theoretical approaches. This may be due to the different choices of the coupling constant and the initial scale.

\begin{table}[hbt!]
  \caption{Comparison of moments of the unpolarized PDF $f_1$ for quark and gluon in our BLFQ approach with other theoretical predictions~\cite{Shi:2022erw, Best:1997qp} at scale $\mu^2=5.76$ GeV$^2$. The uncertainity in our predicted moments is due to the spread in the initial scale $\mu_0^2=0.34\pm 0.034$ GeV$^2$.}
  \label{tab:moments-f1}
  \vspace{0.2cm}
\footnotesize
  \centering
  \begin{tabular}{ l c c c }
    \hline \hline \\
    [-2em]
   & $\langle x \rangle_{f_1^q}$ & $\langle x^2 \rangle_{f_1^q}$ &  $\langle x^3 \rangle_{f_1^q}$  \\
 \\
 [-2em]    
 \hline \\
 [-2em]
 BLFQ (This work) & 0.235 $\pm$ 0.011 & 0.094 $\pm$ 0.007 & 0.045 $\pm$ 0.004\\
 BSE~\cite{Shi:2022erw} & 0.316 & 0.155 & 0.091  \\
 Lattice QCD~\cite{Best:1997qp} & 0.334(21) & 0.174(47) & 0.066(39) \\
\hline
 & $\langle x \rangle_{f_1^g}$ & $\langle x^2 \rangle_{f_1^g}$ &  $\langle x^3 \rangle_{f_1^g}$ \\
  \hline
   BLFQ (This work) & 0.425 $\pm$ 0.013 & 0.063 $\pm$ 0.002 & 0.018 $\pm$ 0.002 \\
   \hline \hline
    \end{tabular}

\end{table}
\begin{table}[hbt!]
  \caption{Comparison of moments of the helicity PDF $g_{1L}$ for quark in our BLFQ approach with other theoretical predictions~\cite{Shi:2022erw, Best:1997qp} at scale $\mu^2=5.76$ GeV$^2$. The uncertainity in our predicted moments is due to the spread in the initial scale $\mu_0^2=0.34\pm 0.034$ GeV$^2$.}
  \vspace{0.2cm}
  \label{tab:moments-g1L}
\footnotesize
  \centering
  \begin{tabular}{ l c c c }
    \hline \hline \\
    [-2em]
 & $\langle x^0 \rangle_{g_{1L}^q}$ & $\langle x \rangle_{g_{1L}^q}$ &  $\langle x^2 \rangle_{g_{1L}^q}$ \\
 \hline
 BLFQ (This work) &  0.476 $\pm$ 0.001 & 0.205 $\pm$ 0.009 & 0.092 $\pm$ 0.006 \\
 BSE~\cite{Shi:2022erw} & 0.660 & 0.227 & 0.111  \\
 Lattice QCD~\cite{Best:1997qp} & 0.570(32) & 0.212(17) & 0.077(34) \\
\hline \hline 
    \end{tabular}
\end{table}



\section{Conclusions}
In this work, we have calculated the quark and gluon distributions of the $\rho$-meson from its LFWFs within the basis light-front quantization (BLFQ) framework. These wave functions have been obtained from the eigen solutions of the light-front QCD Hamiltonian in the light-cone gauge for the light mesons by taking them  into account within $|q\bar{q}\rangle$ and $|q\bar{q}g\rangle$ Fock spaces, together with a 3-d confinement in the
leading Fock sector. In
this study, the gauge link has been set to unity that leaves us only with the T-even quark TMDs.
%
%
We have observed a distinctive feature in the quark TMDs at the low-$x$ domain, which has not been observed in other theoretical approaches for the $\rho$-meson. We have explicitly
shown that the low-$x$ peaked contribution to the quark
TMDs is directly associated with the  inclusion of the dynamical gluon and the related interactions. For $x \gtrsim 0.2$, the quark TMDs in our BLFQ approach have been found to be qualitatively consistent with other theoretical predictions~\cite{Ninomiya:2017ggn, Kaur:2020emh, Shi:2022erw}. Our results  satisfy the positivity bounds implied on the TMDs of spin-1 target~\cite{Bacchetta:2001rb}.

Subsequently, integrating over the valence quark's transverse momentum has yielded four PDFs: $f^q_1(x), g^q_{1L}(x), h^q_1(x)$ and $f^q_{1LL}(x)$. These PDFs exhibit qualitative agreement with the predictions found in the literature~\cite{Ninomiya:2017ggn, Kaur:2020emh, Shi:2022erw}, with the exception of the region $x\lesssim 0.2$. Given the dissimilar behavior of $f_{1LL}$ observed in different models~\cite{Sun:2017gtz, Kaur:2020emh, Ninomiya:2017ggn, Mankiewicz:1988dk, Shi:2022erw}, further investigation and systematic measurements in the future may provide insights into the actual behavior of this PDF. 

In addition, we have explored the gluon PDFs: $f^g_1(x), g^g_{1L}(x), f^g_{1LL}(x)$ and $h^g_{1TT}(x)$. The transversity distribution $h^g_{1TT}$ is near zero in our current treatment of the $\rho$-meson. This is due to the weak $D$-wave component, which is the only contributing factor to this distribution in our calculations. Our future plans involve incorporating sea quarks and additional gluons in our approach, which may result in a nonvanishing gluon transversity distribution. This particular function is crucial as it can only be observed for spin$\geq 1$ systems~\cite{Jaffe:1989xy, Mulders:2000sh, Kumano:2022cdj}. We have further evolved the unpolarized and the helicity-dependent PDFs to the higher scale.                    
The moments of the valence quark and gluon PDFs $f_1$ and $g_{1L}$ have also been evaluated, and compared with the predictions from the Bethe-Salpeter Equation~\cite{Shi:2022erw} and the lattice QCD~\cite{Best:1997qp}. Our ongoing efforts to incorporate additional partons, for example, sea quarks may provide a more comprehensive picture of the $\rho$-meson in momentum space. 


\section*{Acknowledgements}
We thank Siqi Xu, Kaiyu Fu and Sreeraj Nair for many useful discussions. SK is supported by Research Fund for International Young Scientists, Grant No.~12250410251, from the National Natural Science Foundation of China (NSFC), and China Postdoctoral Science Foundation (CPSF), Grant No.~E339951SR0. JL is supported by the Special Research Assistant Funding Project, Chinese Academy of Sciences, by the National Science Foundation of Gansu Province, China, Grant No.~23JRRA631, and National Natural Science Foundation of China under Grant No. 12305095. CM is supported by new faculty start up funding by the Institute of Modern Physics, Chinese Academy of Sciences, Grant No.~E129952YR0. XZ is supported by new faculty startup funding by the Institute of Modern Physics, Chinese Academy of Sciences, by Key Research Program of Frontier Sciences, Chinese Academy of Sciences, Grant No.~ZDBS-LY-7020, by the Natural Science Foundation of Gansu Province, China, Grant No.~20JR10RA067, by the Foundation for Key Talents of Gansu Province, by the Central Funds Guiding the Local Science and Technology Development of Gansu Province, Grant No.~22ZY1QA006, by international partnership program of the Chinese Academy of Sciences, Grant No.~016GJHZ2022103FN, by the Strategic Priority Research Program of the Chinese Academy of Sciences, Grant No.~XDB34000000, and by the National Natural Science Foundation of China under Grant No.12375143. JPV acknowledges partial support from the Department of Energy under Grant No.~DE-SC0023692. This research used resources of the National Energy Research Scientific Computing Center (NERSC), a U.S. Department of Energy Office of Science User Facility located at Lawrence Berkeley National Laboratory, operated under Contract No.~DE-AC02-05CH11231 using NERSC award NP-ERCAP0020944. A portion of the computational resources were also provided by Gansu Computing Center. This research is supported by Gansu International Collaboration and Talents Recruitment Base of Particle Physics, and the International Partnership Program of Chinese Academy of Sciences, Grant No.~016GJHZ2022103FN.


\biboptions{sort&compress}
\bibliographystyle{elsarticle-num}
 \bibliographystyle{apsrev4-1}
\bibliography{ref.bib}

\begin{thebibliography}{10}
\expandafter\ifx\csname url\endcsname\relax
  \def\url#1{\texttt{#1}}\fi
\expandafter\ifx\csname urlprefix\endcsname\relax\def\urlprefix{URL }\fi
\expandafter\ifx\csname href\endcsname\relax
  \def\href#1#2{#2} \def\path#1{#1}\fi

\bibitem{Collins:2011ca}
J.~Collins, {New definition of TMD parton densities}, Int. J. Mod. Phys. Conf.
  Ser. 4 (2011) 85--96.
\newblock \href {http://arxiv.org/abs/1107.4123} {\path{arXiv:1107.4123}},
  \href {https://doi.org/10.1142/S2010194511001590}
  {\path{doi:10.1142/S2010194511001590}}.

\bibitem{Angeles-Martinez:2015sea}
R.~Angeles-Martinez, et~al., {Transverse Momentum Dependent (TMD) parton
  distribution functions: status and prospects}, Acta Phys. Polon. B 46~(12)
  (2015) 2501--2534.
\newblock \href {http://arxiv.org/abs/1507.05267} {\path{arXiv:1507.05267}},
  \href {https://doi.org/10.5506/APhysPolB.46.2501}
  {\path{doi:10.5506/APhysPolB.46.2501}}.

\bibitem{Diehl:2015uka}
M.~Diehl, {Introduction to GPDs and TMDs}, Eur. Phys. J. A 52~(6) (2016) 149.
\newblock \href {http://arxiv.org/abs/1512.01328} {\path{arXiv:1512.01328}},
  \href {https://doi.org/10.1140/epja/i2016-16149-3}
  {\path{doi:10.1140/epja/i2016-16149-3}}.

\bibitem{Collins:1981uw}
J.~C. Collins, D.~E. Soper, {Parton Distribution and Decay Functions}, Nucl.
  Phys. B 194 (1982) 445--492.
\newblock \href {https://doi.org/10.1016/0550-3213(82)90021-9}
  {\path{doi:10.1016/0550-3213(82)90021-9}}.

\bibitem{Gluck:1994uf}
M.~Gluck, E.~Reya, A.~Vogt, {Dynamical parton distributions of the proton and
  small x physics}, Z. Phys. C 67 (1995) 433--448.
\newblock \href {https://doi.org/10.1007/BF01624586}
  {\path{doi:10.1007/BF01624586}}.

\bibitem{Martin:1998sq}
A.~D. Martin, R.~G. Roberts, W.~J. Stirling, R.~S. Thorne, {Parton
  distributions: A New global analysis}, Eur. Phys. J. C 4 (1998) 463--496.
\newblock \href {http://arxiv.org/abs/hep-ph/9803445}
  {\path{arXiv:hep-ph/9803445}}, \href {https://doi.org/10.1007/s100520050220}
  {\path{doi:10.1007/s100520050220}}.

\bibitem{Gluck:1998xa}
M.~Gl\"uck, E.~Reya, A.~Vogt, {Dynamical parton distributions revisited}, Eur.
  Phys. J. C 5 (1998) 461--470.
\newblock \href {http://arxiv.org/abs/hep-ph/9806404}
  {\path{arXiv:hep-ph/9806404}}, \href {https://doi.org/10.1007/s100520050289}
  {\path{doi:10.1007/s100520050289}}.

\bibitem{Ralston:1979ys}
J.~P. Ralston, D.~E. Soper, {Production of Dimuons from High-Energy Polarized
  Proton Proton Collisions}, Nucl. Phys. B 152 (1979) 109.
\newblock \href {https://doi.org/10.1016/0550-3213(79)90082-8}
  {\path{doi:10.1016/0550-3213(79)90082-8}}.

\bibitem{Donohue:1980tn}
J.~T. Donohue, S.~A. Gottlieb, {Dilepton Production from Collisions of
  Polarized Spin 1/2 Hadrons: I. General Kinematic Analysis}, Phys. Rev. D 23
  (1981) 2577--2580.
\newblock \href {https://doi.org/10.1103/PhysRevD.23.2577}
  {\path{doi:10.1103/PhysRevD.23.2577}}.

\bibitem{Collins:1984kg}
J.~C. Collins, D.~E. Soper, G.~F. Sterman, {Transverse Momentum Distribution in
  Drell-Yan Pair and W and Z Boson Production}, Nucl. Phys. B 250 (1985)
  199--224.
\newblock \href {https://doi.org/10.1016/0550-3213(85)90479-1}
  {\path{doi:10.1016/0550-3213(85)90479-1}}.

\bibitem{Tangerman:1994eh}
R.~D. Tangerman, P.~J. Mulders, {Intrinsic transverse momentum and the
  polarized Drell-Yan process}, Phys. Rev. D 51 (1995) 3357--3372.
\newblock \href {http://arxiv.org/abs/hep-ph/9403227}
  {\path{arXiv:hep-ph/9403227}}, \href
  {https://doi.org/10.1103/PhysRevD.51.3357}
  {\path{doi:10.1103/PhysRevD.51.3357}}.

\bibitem{Zhou:2009jm}
J.~Zhou, F.~Yuan, Z.-T. Liang, {Transverse momentum dependent quark
  distributions and polarized Drell-Yan processes}, Phys. Rev. D 81 (2010)
  054008.
\newblock \href {http://arxiv.org/abs/0909.2238} {\path{arXiv:0909.2238}},
  \href {https://doi.org/10.1103/PhysRevD.81.054008}
  {\path{doi:10.1103/PhysRevD.81.054008}}.

\bibitem{Collins:2002kn}
J.~C. Collins, {Leading twist single transverse-spin asymmetries: Drell-Yan and
  deep inelastic scattering}, Phys. Lett. B 536 (2002) 43--48.
\newblock \href {http://arxiv.org/abs/hep-ph/0204004}
  {\path{arXiv:hep-ph/0204004}}, \href
  {https://doi.org/10.1016/S0370-2693(02)01819-1}
  {\path{doi:10.1016/S0370-2693(02)01819-1}}.

\bibitem{Brodsky:2002cx}
S.~J. Brodsky, D.~S. Hwang, I.~Schmidt, {Final state interactions and single
  spin asymmetries in semiinclusive deep inelastic scattering}, Phys. Lett. B
  530 (2002) 99--107.
\newblock \href {http://arxiv.org/abs/hep-ph/0201296}
  {\path{arXiv:hep-ph/0201296}}, \href
  {https://doi.org/10.1016/S0370-2693(02)01320-5}
  {\path{doi:10.1016/S0370-2693(02)01320-5}}.

\bibitem{Ji:2004wu}
X.-d. Ji, J.-p. Ma, F.~Yuan, {QCD factorization for semi-inclusive
  deep-inelastic scattering at low transverse momentum}, Phys. Rev. D 71 (2005)
  034005.
\newblock \href {http://arxiv.org/abs/hep-ph/0404183}
  {\path{arXiv:hep-ph/0404183}}, \href
  {https://doi.org/10.1103/PhysRevD.71.034005}
  {\path{doi:10.1103/PhysRevD.71.034005}}.

\bibitem{Bacchetta:2017gcc}
A.~Bacchetta, F.~Delcarro, C.~Pisano, M.~Radici, A.~Signori, {Extraction of
  partonic transverse momentum distributions from semi-inclusive deep-inelastic
  scattering, Drell-Yan and Z-boson production}, JHEP 06 (2017) 081, [Erratum:
  JHEP 06, 051 (2019)].
\newblock \href {http://arxiv.org/abs/1703.10157} {\path{arXiv:1703.10157}},
  \href {https://doi.org/10.1007/JHEP06(2017)081}
  {\path{doi:10.1007/JHEP06(2017)081}}.

\bibitem{Ninomiya:2017ggn}
Y.~Ninomiya, W.~Bentz, I.~C. Clo\"et, {Transverse-momentum-dependent quark
  distribution functions of spin-one targets: Formalism and covariant
  calculations}, Phys. Rev. C 96~(4) (2017) 045206.
\newblock \href {http://arxiv.org/abs/1707.03787} {\path{arXiv:1707.03787}},
  \href {https://doi.org/10.1103/PhysRevC.96.045206}
  {\path{doi:10.1103/PhysRevC.96.045206}}.

\bibitem{Kaur:2020emh}
S.~Kaur, C.~Mondal, H.~Dahiya, {Light-front holographic $\rho$-meson
  distributions in the momentum space}, JHEP 01 (2021) 136.
\newblock \href {http://arxiv.org/abs/2009.04288} {\path{arXiv:2009.04288}},
  \href {https://doi.org/10.1007/JHEP01(2021)136}
  {\path{doi:10.1007/JHEP01(2021)136}}.

\bibitem{Shi:2022erw}
C.~Shi, J.~Li, M.~Li, X.~Chen, W.~Jia, {Transverse momentum distributions of
  valence quarks in light and heavy vector mesons}, Phys. Rev. D 106~(1) (2022)
  014026.
\newblock \href {http://arxiv.org/abs/2205.02757} {\path{arXiv:2205.02757}},
  \href {https://doi.org/10.1103/PhysRevD.106.014026}
  {\path{doi:10.1103/PhysRevD.106.014026}}.

\bibitem{Best:1997qp}
C.~Best, M.~Gockeler, R.~Horsley, E.-M. Ilgenfritz, H.~Perlt, P.~E.~L. Rakow,
  A.~Schafer, G.~Schierholz, A.~Schiller, S.~Schramm, {Pion and rho structure
  functions from lattice QCD}, Phys. Rev. D 56 (1997) 2743--2754.
\newblock \href {http://arxiv.org/abs/hep-lat/9703014}
  {\path{arXiv:hep-lat/9703014}}, \href
  {https://doi.org/10.1103/PhysRevD.56.2743}
  {\path{doi:10.1103/PhysRevD.56.2743}}.

\bibitem{Loffler:2021afv}
M.~L\"offler, P.~Wein, T.~Wurm, S.~Weish\"aupl, D.~Jenkins, R.~R\"odl,
  A.~Sch\"afer, L.~Walter, {Mellin moments of spin dependent and independent
  PDFs of the pion and rho meson}, Phys. Rev. D 105~(1) (2022) 014505.
\newblock \href {http://arxiv.org/abs/2108.07544} {\path{arXiv:2108.07544}},
  \href {https://doi.org/10.1103/PhysRevD.105.014505}
  {\path{doi:10.1103/PhysRevD.105.014505}}.

\bibitem{Bacchetta:2000jk}
A.~Bacchetta, P.~J. Mulders, {Deep inelastic leptoproduction of spin-one
  hadrons}, Phys. Rev. D 62 (2000) 114004.
\newblock \href {http://arxiv.org/abs/hep-ph/0007120}
  {\path{arXiv:hep-ph/0007120}}, \href
  {https://doi.org/10.1103/PhysRevD.62.114004}
  {\path{doi:10.1103/PhysRevD.62.114004}}.

\bibitem{Bacchetta:2001rb}
A.~Bacchetta, P.~J. Mulders, {Positivity bounds on spin one distribution and
  fragmentation functions}, Phys. Lett. B 518 (2001) 85--93.
\newblock \href {http://arxiv.org/abs/hep-ph/0104176}
  {\path{arXiv:hep-ph/0104176}}, \href
  {https://doi.org/10.1016/S0370-2693(01)01051-6}
  {\path{doi:10.1016/S0370-2693(01)01051-6}}.

\bibitem{Hino:1998ww}
S.~Hino, S.~Kumano, {Structure functions in the polarized Drell-Yan processes
  with spin 1/2 and spin 1 hadrons. 1. General formalism}, Phys. Rev. D 59
  (1999) 094026.
\newblock \href {http://arxiv.org/abs/hep-ph/9810425}
  {\path{arXiv:hep-ph/9810425}}, \href
  {https://doi.org/10.1103/PhysRevD.59.094026}
  {\path{doi:10.1103/PhysRevD.59.094026}}.

\bibitem{Hino:1999qi}
S.~Hino, S.~Kumano, {Structure functions in the polarized Drell-Yan processes
  with spin 1/2 and spin 1 hadrons. 2. Parton model}, Phys. Rev. D 60 (1999)
  054018.
\newblock \href {http://arxiv.org/abs/hep-ph/9902258}
  {\path{arXiv:hep-ph/9902258}}, \href
  {https://doi.org/10.1103/PhysRevD.60.054018}
  {\path{doi:10.1103/PhysRevD.60.054018}}.

\bibitem{Hoodbhoy:1988am}
P.~Hoodbhoy, R.~L. Jaffe, A.~Manohar, {Novel Effects in Deep Inelastic
  Scattering from Spin 1 Hadrons}, Nucl. Phys. B 312 (1989) 571--588.
\newblock \href {https://doi.org/10.1016/0550-3213(89)90572-5}
  {\path{doi:10.1016/0550-3213(89)90572-5}}.

\bibitem{Close:1990zw}
F.~E. Close, S.~Kumano, {A sum rule for the spin dependent structure function
  b-1(x) for spin one hadrons}, Phys. Rev. D 42 (1990) 2377--2379.
\newblock \href {https://doi.org/10.1103/PhysRevD.42.2377}
  {\path{doi:10.1103/PhysRevD.42.2377}}.

\bibitem{Umnikov:1996qv}
A.~Y. Umnikov, {Relativistic calculation of structure functions b(1,2)(x) of
  the deuteron}, Phys. Lett. B 391 (1997) 177--184.
\newblock \href {http://arxiv.org/abs/hep-ph/9605291}
  {\path{arXiv:hep-ph/9605291}}, \href
  {https://doi.org/10.1016/S0370-2693(96)01440-2}
  {\path{doi:10.1016/S0370-2693(96)01440-2}}.

\bibitem{Detmold:2005iz}
W.~Detmold, {Target mass effects in deep-inelastic scattering on the deuteron},
  Phys. Lett. B 632 (2006) 261--269.
\newblock \href {http://arxiv.org/abs/hep-ph/0509011}
  {\path{arXiv:hep-ph/0509011}}, \href
  {https://doi.org/10.1016/j.physletb.2005.10.091}
  {\path{doi:10.1016/j.physletb.2005.10.091}}.

\bibitem{Kumano:2010vz}
S.~Kumano, {Tensor-polarized quark and antiquark distribution functions in a
  spin-one hadron}, Phys. Rev. D 82 (2010) 017501.
\newblock \href {http://arxiv.org/abs/1005.4524} {\path{arXiv:1005.4524}},
  \href {https://doi.org/10.1103/PhysRevD.82.017501}
  {\path{doi:10.1103/PhysRevD.82.017501}}.

\bibitem{Islam:2012zua}
S.~Islam, D.~K. Choudhury, {Tensor structure function b1(d)(x, Q**2) of the
  deuteron at NLO and NNLO at small-x}, Eur. Phys. J. C 72 (2012) 2257.
\newblock \href {https://doi.org/10.1140/epjc/s10052-012-2257-x}
  {\path{doi:10.1140/epjc/s10052-012-2257-x}}.

\bibitem{Miller:2013hla}
G.~A. Miller, {Pionic and Hidden-Color, Six-Quark Contributions to the Deuteron
  b1 Structure Function}, Phys. Rev. C 89~(4) (2014) 045203.
\newblock \href {http://arxiv.org/abs/1311.4561} {\path{arXiv:1311.4561}},
  \href {https://doi.org/10.1103/PhysRevC.89.045203}
  {\path{doi:10.1103/PhysRevC.89.045203}}.

\bibitem{Boer:2016xqr}
D.~Boer, S.~Cotogno, T.~van Daal, P.~J. Mulders, A.~Signori, Y.-J. Zhou, {Gluon
  and Wilson loop TMDs for hadrons of spin $\leq$ 1}, JHEP 10 (2016) 013.
\newblock \href {http://arxiv.org/abs/1607.01654} {\path{arXiv:1607.01654}},
  \href {https://doi.org/10.1007/JHEP10(2016)013}
  {\path{doi:10.1007/JHEP10(2016)013}}.

\bibitem{Cotogno:2017puy}
S.~Cotogno, T.~van Daal, P.~J. Mulders, {Positivity bounds on gluon TMDs for
  hadrons of spin $\le$ 1}, JHEP 11 (2017) 185.
\newblock \href {http://arxiv.org/abs/1709.07827} {\path{arXiv:1709.07827}},
  \href {https://doi.org/10.1007/JHEP11(2017)185}
  {\path{doi:10.1007/JHEP11(2017)185}}.

\bibitem{Shi:2023oll}
C.~Shi, J.~Li, P.-L. Yin, W.~Jia, {Unpolarized generalized parton distributions
  of light and heavy vector mesons}, Phys. Rev. D 107~(7) (2023) 074009.
\newblock \href {http://arxiv.org/abs/2302.02388} {\path{arXiv:2302.02388}},
  \href {https://doi.org/10.1103/PhysRevD.107.074009}
  {\path{doi:10.1103/PhysRevD.107.074009}}.

\bibitem{Vary:2009gt}
J.~P. Vary, H.~Honkanen, J.~Li, P.~Maris, S.~J. Brodsky, A.~Harindranath, G.~F.
  de~Teramond, P.~Sternberg, E.~G. Ng, C.~Yang, {Hamiltonian light-front field
  theory in a basis function approach}, Phys. Rev. C 81 (2010) 035205.
\newblock \href {http://arxiv.org/abs/0905.1411} {\path{arXiv:0905.1411}},
  \href {https://doi.org/10.1103/PhysRevC.81.035205}
  {\path{doi:10.1103/PhysRevC.81.035205}}.

\bibitem{Lan:2019vui}
J.~Lan, C.~Mondal, S.~Jia, X.~Zhao, J.~P. Vary, {Parton Distribution Functions
  from a Light Front Hamiltonian and QCD Evolution for Light Mesons}, Phys.
  Rev. Lett. 122~(17) (2019) 172001.
\newblock \href {http://arxiv.org/abs/1901.11430} {\path{arXiv:1901.11430}},
  \href {https://doi.org/10.1103/PhysRevLett.122.172001}
  {\path{doi:10.1103/PhysRevLett.122.172001}}.

\bibitem{Lan:2019rba}
J.~Lan, C.~Mondal, S.~Jia, X.~Zhao, J.~P. Vary, {Pion and kaon parton
  distribution functions from basis light front quantization and QCD
  evolution}, Phys. Rev. D 101~(3) (2020) 034024.
\newblock \href {http://arxiv.org/abs/1907.01509} {\path{arXiv:1907.01509}},
  \href {https://doi.org/10.1103/PhysRevD.101.034024}
  {\path{doi:10.1103/PhysRevD.101.034024}}.

\bibitem{Mondal:2019jdg}
C.~Mondal, S.~Xu, J.~Lan, X.~Zhao, Y.~Li, D.~Chakrabarti, J.~P. Vary, {Proton
  structure from a light-front Hamiltonian}, Phys. Rev. D 102~(1) (2020)
  016008.
\newblock \href {http://arxiv.org/abs/1911.10913} {\path{arXiv:1911.10913}},
  \href {https://doi.org/10.1103/PhysRevD.102.016008}
  {\path{doi:10.1103/PhysRevD.102.016008}}.

\bibitem{Xu:2021wwj}
S.~Xu, C.~Mondal, J.~Lan, X.~Zhao, Y.~Li, J.~P. Vary, {Nucleon structure from
  basis light-front quantization}, Phys. Rev. D 104~(9) (2021) 094036.
\newblock \href {http://arxiv.org/abs/2108.03909} {\path{arXiv:2108.03909}},
  \href {https://doi.org/10.1103/PhysRevD.104.094036}
  {\path{doi:10.1103/PhysRevD.104.094036}}.

\bibitem{Mondal:2021czk}
C.~Mondal, S.~Nair, S.~Jia, X.~Zhao, J.~P. Vary, {Pion to photon transition
  form factors with basis light-front quantization}, Phys. Rev. D 104~(9)
  (2021) 094034.
\newblock \href {http://arxiv.org/abs/2109.02279} {\path{arXiv:2109.02279}},
  \href {https://doi.org/10.1103/PhysRevD.104.094034}
  {\path{doi:10.1103/PhysRevD.104.094034}}.

\bibitem{Lan:2021wok}
J.~Lan, K.~Fu, C.~Mondal, X.~Zhao, j.~P. Vary, {Light mesons with one dynamical
  gluon on the light front}, Phys. Lett. B 825 (2022) 136890.
\newblock \href {http://arxiv.org/abs/2106.04954} {\path{arXiv:2106.04954}},
  \href {https://doi.org/10.1016/j.physletb.2022.136890}
  {\path{doi:10.1016/j.physletb.2022.136890}}.

\bibitem{Liu:2022fvl}
Y.~Liu, S.~Xu, C.~Mondal, X.~Zhao, J.~P. Vary, {Angular momentum and
  generalized parton distributions for the proton with basis light-front
  quantization}, Phys. Rev. D 105~(9) (2022) 094018.
\newblock \href {http://arxiv.org/abs/2202.00985} {\path{arXiv:2202.00985}},
  \href {https://doi.org/10.1103/PhysRevD.105.094018}
  {\path{doi:10.1103/PhysRevD.105.094018}}.

\bibitem{Hu:2022ctr}
Z.~Hu, S.~Xu, C.~Mondal, X.~Zhao, J.~P. Vary, {Transverse momentum structure of
  proton within the basis light-front quantization framework}, Phys. Lett. B
  833 (2022) 137360.
\newblock \href {http://arxiv.org/abs/2205.04714} {\path{arXiv:2205.04714}},
  \href {https://doi.org/10.1016/j.physletb.2022.137360}
  {\path{doi:10.1016/j.physletb.2022.137360}}.

\bibitem{Peng:2022lte}
T.~Peng, Z.~Zhu, S.~Xu, X.~Liu, C.~Mondal, X.~Zhao, J.~P. Vary, {Basis
  light-front quantization approach to \ensuremath{\Lambda} and
  \ensuremath{\Lambda}c and their isospin triplet baryons}, Phys. Rev. D
  106~(11) (2022) 114040.
\newblock \href {http://arxiv.org/abs/2208.00355} {\path{arXiv:2208.00355}},
  \href {https://doi.org/10.1103/PhysRevD.106.114040}
  {\path{doi:10.1103/PhysRevD.106.114040}}.

\bibitem{Xu:2022abw}
S.~Xu, C.~Mondal, X.~Zhao, Y.~Li, J.~P. Vary, {Nucleon spin decomposition with
  one dynamical gluon} (9 2022).
\newblock \href {http://arxiv.org/abs/2209.08584} {\path{arXiv:2209.08584}}.

\bibitem{Zhu:2023lst}
Z.~Zhu, Z.~Hu, J.~Lan, C.~Mondal, X.~Zhao, J.~P. Vary, {Transverse structure of
  the pion beyond leading twist with basis light-front quantization}, Phys.
  Lett. B 839 (2023) 137808.
\newblock \href {http://arxiv.org/abs/2301.12994} {\path{arXiv:2301.12994}},
  \href {https://doi.org/10.1016/j.physletb.2023.137808}
  {\path{doi:10.1016/j.physletb.2023.137808}}.

\bibitem{Zhu:2023nhl}
Z.~Zhu, T.~Peng, Z.~Hu, S.~Xu, C.~Mondal, X.~Zhao, J.~P. Vary, {Transverse
  momentum structure of strange and charmed baryons: A light-front Hamiltonian
  approach}, Phys. Rev. D 108~(3) (2023) 036009.
\newblock \href {http://arxiv.org/abs/2304.05058} {\path{arXiv:2304.05058}},
  \href {https://doi.org/10.1103/PhysRevD.108.036009}
  {\path{doi:10.1103/PhysRevD.108.036009}}.

\bibitem{Kaur:2023lun}
S.~Kaur, S.~Xu, C.~Mondal, X.~Zhao, J.~P. Vary, {Spatial imaging of proton via
  leading-twist GPDs with basis light-front quantization} (7 2023).
\newblock \href {http://arxiv.org/abs/2307.09869} {\path{arXiv:2307.09869}}.

\bibitem{Lin:2023ezw}
B.~Lin, S.~Nair, S.~Xu, Z.~Hu, C.~Mondal, X.~Zhao, J.~P. Vary, {Generalized
  parton distributions of gluon in proton: a light-front quantization approach}
  (8 2023).
\newblock \href {http://arxiv.org/abs/2308.08275} {\path{arXiv:2308.08275}}.

\bibitem{Brodsky:1997de}
S.~J. Brodsky, H.-C. Pauli, S.~S. Pinsky, {Quantum chromodynamics and other
  field theories on the light cone}, Phys. Rept. 301 (1998) 299--486.
\newblock \href {http://arxiv.org/abs/hep-ph/9705477}
  {\path{arXiv:hep-ph/9705477}}, \href
  {https://doi.org/10.1016/S0370-1573(97)00089-6}
  {\path{doi:10.1016/S0370-1573(97)00089-6}}.

\bibitem{Glazek:1993rc}
S.~D. Glazek, K.~G. Wilson, {Renormalization of Hamiltonians}, Phys. Rev. D 48
  (1993) 5863--5872.
\newblock \href {https://doi.org/10.1103/PhysRevD.48.5863}
  {\path{doi:10.1103/PhysRevD.48.5863}}.

\bibitem{Karmanov:2008br}
V.~A. Karmanov, J.~F. Mathiot, A.~V. Smirnov, {Systematic renormalization
  scheme in light-front dynamics with Fock space truncation}, Phys. Rev. D 77
  (2008) 085028.
\newblock \href {http://arxiv.org/abs/0801.4507} {\path{arXiv:0801.4507}},
  \href {https://doi.org/10.1103/PhysRevD.77.085028}
  {\path{doi:10.1103/PhysRevD.77.085028}}.

\bibitem{Karmanov:2012aj}
V.~A. Karmanov, J.~F. Mathiot, A.~V. Smirnov, {Ab initio nonperturbative
  calculation of physical observables in light-front dynamics. Application to
  the Yukawa model}, Phys. Rev. D 86 (2012) 085006.
\newblock \href {http://arxiv.org/abs/1204.3257} {\path{arXiv:1204.3257}},
  \href {https://doi.org/10.1103/PhysRevD.86.085006}
  {\path{doi:10.1103/PhysRevD.86.085006}}.

\bibitem{Zhao:2014hpa}
X.~Zhao, {Advances in Basis Light-front Quantization}, Few Body Syst. 56~(6-9)
  (2015) 257--265.
\newblock \href {http://arxiv.org/abs/1411.7748} {\path{arXiv:1411.7748}},
  \href {https://doi.org/10.1007/s00601-015-1003-y}
  {\path{doi:10.1007/s00601-015-1003-y}}.

\bibitem{Zhao:2020kuf}
X.~Zhao, K.~Fu, H.~Zhao, J.~P. Vary, {Positronium: an illustration of
  nonperturbative renormalization in a basis light-front approach}, PoS LC2019
  (2020) 090.
\newblock \href {http://arxiv.org/abs/2103.06719} {\path{arXiv:2103.06719}},
  \href {https://doi.org/10.22323/1.374.0090} {\path{doi:10.22323/1.374.0090}}.

\bibitem{Glazek:1992aq}
S.~D. Glazek, R.~J. Perry, {Special example of relativistic Hamiltonian field
  theory}, Phys. Rev. D 45 (1992) 3740--3754.
\newblock \href {https://doi.org/10.1103/PhysRevD.45.3740}
  {\path{doi:10.1103/PhysRevD.45.3740}}.

\bibitem{Burkardt:1998dd}
M.~Burkardt, {Dynamical vertex mass generation and chiral symmetry breaking on
  the light front}, Phys. Rev. D 58 (1998) 096015.
\newblock \href {http://arxiv.org/abs/hep-th/9805088}
  {\path{arXiv:hep-th/9805088}}, \href
  {https://doi.org/10.1103/PhysRevD.58.096015}
  {\path{doi:10.1103/PhysRevD.58.096015}}.

\bibitem{Li:2015zda}
Y.~Li, P.~Maris, X.~Zhao, J.~P. Vary, {Heavy Quarkonium in a Holographic
  Basis}, Phys. Lett. B 758 (2016) 118--124.
\newblock \href {http://arxiv.org/abs/1509.07212} {\path{arXiv:1509.07212}},
  \href {https://doi.org/10.1016/j.physletb.2016.04.065}
  {\path{doi:10.1016/j.physletb.2016.04.065}}.

\bibitem{Xu:2023nqv}
S.~Xu, C.~Mondal, X.~Zhao, Y.~Li, J.~P. Vary, {Quark and gluon spin and orbital
  angular momentum in the proton}, Phys. Rev. D 108~(9) (2023) 094002.
\newblock \href {https://doi.org/10.1103/PhysRevD.108.094002}
  {\path{doi:10.1103/PhysRevD.108.094002}}.

\bibitem{Brodsky:2014yha}
S.~J. Brodsky, G.~F. de~Teramond, H.~G. Dosch, J.~Erlich, {Light-Front
  Holographic QCD and Emerging Confinement}, Phys. Rept. 584 (2015) 1--105.
\newblock \href {http://arxiv.org/abs/1407.8131} {\path{arXiv:1407.8131}},
  \href {https://doi.org/10.1016/j.physrep.2015.05.001}
  {\path{doi:10.1016/j.physrep.2015.05.001}}.

\bibitem{Zhao:2014xaa}
X.~Zhao, H.~Honkanen, P.~Maris, J.~P. Vary, S.~J. Brodsky, {Electron g-2 in
  Light-Front Quantization}, Phys. Lett. B 737 (2014) 65--69.
\newblock \href {http://arxiv.org/abs/1402.4195} {\path{arXiv:1402.4195}},
  \href {https://doi.org/10.1016/j.physletb.2014.08.020}
  {\path{doi:10.1016/j.physletb.2014.08.020}}.

\bibitem{Mulders2001TransverseMD}
P.~Mulders, Transverse momentum dependence in structure functions in hard
  scattering processes, 2001.

\bibitem{Pasquini:2014ppa}
B.~Pasquini, P.~Schweitzer, {Pion transverse momentum dependent parton
  distributions in a light-front constituent approach, and the Boer-Mulders
  effect in the pion-induced Drell-Yan process}, Phys. Rev. D 90~(1) (2014)
  014050.
\newblock \href {http://arxiv.org/abs/1406.2056} {\path{arXiv:1406.2056}},
  \href {https://doi.org/10.1103/PhysRevD.90.014050}
  {\path{doi:10.1103/PhysRevD.90.014050}}.

\bibitem{Noguera:2015iia}
S.~Noguera, S.~Scopetta, {Pion transverse momentum dependent parton
  distributions in the Nambu and Jona-Lasinio model}, JHEP 11 (2015) 102.
\newblock \href {http://arxiv.org/abs/1508.01061} {\path{arXiv:1508.01061}},
  \href {https://doi.org/10.1007/JHEP11(2015)102}
  {\path{doi:10.1007/JHEP11(2015)102}}.

\bibitem{Ahmady:2019yvo}
M.~Ahmady, C.~Mondal, R.~Sandapen, {Predicting the light-front holographic TMDs
  of the pion}, Phys. Rev. D 100~(5) (2019) 054005.
\newblock \href {http://arxiv.org/abs/1907.06561} {\path{arXiv:1907.06561}},
  \href {https://doi.org/10.1103/PhysRevD.100.054005}
  {\path{doi:10.1103/PhysRevD.100.054005}}.

\bibitem{Sun:2017gtz}
B.-D. Sun, Y.-B. Dong, {$\rho$ meson unpolarized generalized parton
  distributions with a light-front constituent quark model}, Phys. Rev. D
  96~(3) (2017) 036019.
\newblock \href {http://arxiv.org/abs/1707.03972} {\path{arXiv:1707.03972}},
  \href {https://doi.org/10.1103/PhysRevD.96.036019}
  {\path{doi:10.1103/PhysRevD.96.036019}}.

\bibitem{Mankiewicz:1988dk}
L.~Mankiewicz, {Simple Relativistic Model for B1 Structure Function of the
  $\rho$ Meson}, Phys. Rev. D 40 (1989) 255.
\newblock \href {https://doi.org/10.1103/PhysRevD.40.255}
  {\path{doi:10.1103/PhysRevD.40.255}}.

\bibitem{Meissner:2007rx}
S.~Meissner, A.~Metz, K.~Goeke, {Relations between generalized and transverse
  momentum dependent parton distributions}, Phys. Rev. D 76 (2007) 034002.
\newblock \href {http://arxiv.org/abs/hep-ph/0703176}
  {\path{arXiv:hep-ph/0703176}}, \href
  {https://doi.org/10.1103/PhysRevD.76.034002}
  {\path{doi:10.1103/PhysRevD.76.034002}}.

\bibitem{Jaffe:1989xy}
R.~L. Jaffe, A.~Manohar, {NUCLEAR GLUONOMETRY}, Phys. Lett. B 223 (1989)
  218--224.
\newblock \href {https://doi.org/10.1016/0370-2693(89)90242-6}
  {\path{doi:10.1016/0370-2693(89)90242-6}}.

\bibitem{Mulders:2000sh}
P.~J. Mulders, J.~Rodrigues, {Transverse momentum dependence in gluon
  distribution and fragmentation functions}, Phys. Rev. D 63 (2001) 094021.
\newblock \href {http://arxiv.org/abs/hep-ph/0009343}
  {\path{arXiv:hep-ph/0009343}}, \href
  {https://doi.org/10.1103/PhysRevD.63.094021}
  {\path{doi:10.1103/PhysRevD.63.094021}}.

\bibitem{Kumano:2022cdj}
S.~Kumano, Q.-T. Song, {Gluon transversity and TMDs for spin-1 hadrons}, Rev.
  Mex. Fis. Suppl. 3~(3) (2022) 0308097.
\newblock \href {http://arxiv.org/abs/2201.04875} {\path{arXiv:2201.04875}},
  \href {https://doi.org/10.31349/SuplRevMexFis.3.0308097}
  {\path{doi:10.31349/SuplRevMexFis.3.0308097}}.

\bibitem{Dokshitzer:1977sg}
Y.~L. Dokshitzer, {Calculation of the Structure Functions for Deep Inelastic
  Scattering and e+ e- Annihilation by Perturbation Theory in Quantum
  Chromodynamics.}, Sov. Phys. JETP 46 (1977) 641--653.

\bibitem{Gribov:1972ri}
V.~N. Gribov, L.~N. Lipatov, {Deep inelastic e p scattering in perturbation
  theory}, Sov. J. Nucl. Phys. 15 (1972) 438--450.

\bibitem{Altarelli:1977zs}
G.~Altarelli, G.~Parisi, {Asymptotic Freedom in Parton Language}, Nucl. Phys. B
  126 (1977) 298--318.
\newblock \href {https://doi.org/10.1016/0550-3213(77)90384-4}
  {\path{doi:10.1016/0550-3213(77)90384-4}}.

\end{thebibliography}
\end{document}